
%
%
%
\input harvmac.tex
\input pictex.tex
\let\includefigures=\iftrue
\includefigures
\input epsf
\epsfclipon
\def\fig#1#2{\topinsert\epsffile{#1}\noindent{#2}\endinsert}
\else
\def\fig#1#2{}
\fi
\def\Title#1#2{\rightline{#1}
\ifx\answ\bigans\nopagenumbers\pageno0\vskip1in%
\baselineskip 15pt plus 1pt minus 1pt
\else
\def\listrefs{\footatend\vskip 1in\immediate\closeout\rfile\writestoppt
\baselineskip=14pt\centerline{{\bf References}}\bigskip{\frenchspacing%
\parindent=20pt\escapechar=` \input
refs.tmp\vfill\eject}\nonfrenchspacing}
\pageno1\vskip.8in\fi \centerline{\titlefont #2}\vskip .5in}

\ifx\answ\bigans\def\tcbreak#1{}\else\def\tcbreak#1{\cr&{#1}}\fi
%

\def\inbar{\,\vrule height1.5ex width.4pt depth0pt}
\def\IB{\relax{\rm I\kern-.18em B}}
\def\IC{\relax\hbox{$\inbar\kern-.3em{\rm C}$}}
\def\IP{\relax{\rm I\kern-.18em P}}
\def\IR{\relax{\rm I\kern-.18em R}}

\def\tr{{\rm Tr}}

\def\sgn{{\rm sgn~}}

\def\D{\Delta}

\def\barint{-\hskip -11pt\int}
\Title{\vbox{\baselineskip12pt
\hfill{\vbox{
\hbox{LPTENS-95/9}}}}}
{\vbox{\centerline{Character Expansion Methods for Matrix Models}
\centerline{ }
\centerline{of
Dually Weighted Graphs}}}
\centerline{Vladimir A. Kazakov}
\smallskip
\centerline{Matthias Staudacher $^\dagger$}
\smallskip
\centerline{{\it and}}
\smallskip
\centerline{Thomas Wynter $^\dagger$}\footnote~{
\hskip -11.5pt $^\dagger$ \hskip 4pt
This work is supported by funds provided by the European Community,
Human Capital and Mobility  Programme.}
\bigskip
\centerline{Laboratoire de Physique Th\'eorique de}
\centerline{l'\'Ecole Normale Sup\'erieure\footnote*{
Unit\'e Propre du
Centre National de la Recherche Scientifique,
associ\'ee \`a l'\'Ecole Normale Sup\'erieure et \`a
l'Universit\'e de Paris-Sud.}}
\bigskip
\noindent
We consider generalized one-matrix models in which external fields
allow control over the coordination numbers on both the original and
dual lattices. We rederive in a simple fashion a character expansion
formula for these models originally due to Itzykson and Di Francesco,
and then demonstrate how to take the large N limit of this expansion.
The relationship to the usual matrix model resolvent is elucidated.
Our methods give as a by-product an extremely simple derivation of the
Migdal integral equation describing the large $N$ limit of the
Itzykson-Zuber formula. We illustrate and check our methods by
analyzing a number of models solvable by traditional means.  We then
proceed to solve a new model: a sum over planar graphs possessing even
coordination numbers on both the original and the dual lattice.  We
conclude by formulating equations for the case of arbitrary sets of
even, self-dual coupling constants. This opens the way for studying
the deep problem of phise transitions from random to flat lattices.
\Date{January 12, 1995}
\nref\KPZ{V.G. Knizhnik, A.M. Polyakov \& A.B. Zamolodchikov, Mod.
Phys. Lett. A3 (1988) 819; F. David, Mod Phys. Lett. A3 (1988) 1651;
H. Kawai, Nucl. Phys. B321 (1989) 509.}
\nref\DS{M.R. Douglas, Phys. Lett. B348 (1991) 507.}
\nref\KaKo{V.A. Kazakov \& I.K. Kostov, Nucl. Phys. B386 (1992) 520.}
\nref\IDiF{P. Di Francesco \& C. Itzykson, Ann. Inst. Henri.
Poincar\'e Vol. 59, no. 2 (1993) 117.}
\nref\VKaz{V.A. Kazakov, Nucl. Phys. B354 (1991) 614.}
\nref\KW{V.A. Kazakov \& T. Wynter, preprint LPTENS-94/28 (1994).}
\nref\DougKaz{M.R. Douglas \& V.A. Kazakov, Phys. Lett. 319B (1993)
219.}
\nref\DDSW{S.R. Das, A. Dhar, A.M. Sengupta \& S.R. Wadia, Mod. Phys.
Lett. A5 (1990) 1041.}
\nref\Mig{A.A. Migdal, Mod. Phys. Lett. A8 No.4 (1993) 359-371.}
\nref\Matyt{A. Matytsin, Nucl. Phys. B411 (1994) 805.}
\nref\BIPZ{E. Brezin, C. Itzykson, G. Parisi \& J.B. Zuber, Commun.
Math. Phys. 59 (1978), 35.}
\nref\mss{G. Moore, N. Seiberg \& M. Staudacher, Nucl. Phys.
B362 (1991) 665.}
\nref\GrossTaylor{D. Gross \& W. Taylor, Nucl. Phys. B 400 (1993) 181.}
\nref\Kawai{H. Kawaii and R. Nakayama, Phys. Lett. B306 (1993) 224.}
\nref\BK{D.V. Boulatov \& V.A. Kazakov, Phys. Lett. B184 (1986) 247.}
\nref\Kaz{V.A. Kazakov, Phys. Lett. 119A (1986) 140.}
\newsec{Introduction}

After the considerable success of two dimensional quantum field theory
and statistical mechanics, - integrable models on 2D regular lattices,
conformal field theories, Liouville theory and matrix models of 2D
gravity and non-critical strings - progress in analytical results in
this field has slowed down.

Among the principal questions remaining unsolved are, first, the so
called $c=1$ barrier for non-critical strings (c is the central charge
of the matter), and, second, the mysterious connection between the
physical properties of various integrable 2D models coupled and
non-coupled to gravity. The first problem is usually atributed to the
absence of a stable vacuum for $c>1$, though it has never been clearly
understood. Indeed, in terms of matrix models, the obstacles seem to
be purely technical. The second problem concerns the observation of
many intriguing relations between 2D physical systems with and without
coupling to 2D quantum gravity.

This relation is clearly established on the level of critical
exponents by the use of the continuous formulation of 2D gravity
\KPZ.  The conformal dimensions of matter fields undergo a simple
quadratic transformation as the result of gravitational dressing. The
same phenomenon is observed, of course, in the matrix model formalism.
However, if one goes away from the critical point, the relation
between the physical properties with and without gravitational
coupling, although persisting, becomes much more tricky and
fragmentary. For example there is the description of 2D gravitational
systems in terms of KDV hierarchies of classical 2D integrable systems
\DS, as well as a strange coincidence between the amplitude for open
string in SOS formalism and the S-matrix of the two dimensional
Sine-Gordon theory \KaKo.

It seems that two-dimensional physics is more united than one would
think at first sight. An interesting question to ask would be the
existence of some interpolating models connecting the gravitational
and ``flat'' phases of the same matter fields.

Our paper is inspired by this physical idea, though it concerns mostly
the elaboration of a technique for the solution of a new type of
matrix model. The model describes, in the large N limit, planar graphs
having arbitrary coordination number dependent weights for both the
vertices and faces. In other words, we introduce a set of couplings
$t_1,t_2,...t_q,...$, the weights of vertices with $1,2,...,q,...$
neighbours, and $t_1^*,t_2^*,...t_q^*,...$, the weights of the dual
vertices (or faces) with appropriate coordination numbers.  In Fig. 1
below is a typical graph with, for illustration, some vertices on the
original and dual lattice labeled with their associated weights.  The
matrix models under consideration allow us to generate the following
partition function of closed planar graphs $G$:
\eqn\DWG{
Z(t,t^*)=
\sum_{G} \prod_{v_q,v_q^* \in G} t_q^{\# v_q}\ {t_q^{*}}^{ \# v_q^*}}
where $v_q,v_q^*$ are the vertices with $q$ neighbours on the original
and dual graph, respectively, and $\# v_q,\# v_q^*$ are the numbers of
such vertices in the given graph $G$. We propose to call this the
model of dually weighted graphs (DWG).
\vskip 20pt
\beginpicture

\setcoordinatesystem units < 1.000cm, 1.000cm>
\linethickness=1pt
\setdashes < 0.1270cm>
\plot 14.446 13.557 15.272 12.383 /
\plot 14.161 11.748 15.272 12.256 /
\plot 15.050 10.351 16.510 11.239 /
\plot 15.272 12.256 16.510 11.239 /
\plot 15.399 12.319 16.066 13.049 /
\plot 16.066 13.049 16.066 13.049 /
\plot 16.066 13.049 16.732 12.605 /
\plot 16.732 12.605 16.574 11.366 /
\plot 16.859 12.605 16.732 11.430 /
\plot 16.859 11.303 18.034 11.811 /
\plot 15.113 10.192 16.574 11.081 /
\plot 16.574 11.017 16.447  9.684 /
\plot 16.732 11.081 16.574  9.684 /
\plot 16.732 11.081 18.193 10.414 /
\plot 16.796 11.239 18.256 10.573 /
\plot 16.732 11.366 17.971 11.970 /
\plot 16.859 12.605 17.685 12.986 /
\plot 16.129 13.208 16.796 12.764 /
\plot 16.796 12.764 17.590 13.113 /
\plot 17.590 13.113 17.685 14.383 /
\plot 17.748 13.113 17.812 14.287 /
\plot 17.748 13.113 18.193 13.049 /
\plot 17.812 12.986 18.129 12.922 /
\plot 16.129 13.208 16.351 14.224 /
\plot 16.351 14.224 17.685 14.383 /
\plot 16.351 14.383 17.590 14.510 /
\plot 16.351 14.383 16.288 15.176 /
\plot 16.288 15.240 17.399 14.796 /
\plot 16.447 15.304 17.399 14.954 /
\plot 16.510 15.399 17.399 15.684 /
\plot 14.605 13.652 15.399 12.478 /
\plot 15.399 12.541 15.907 13.113 /
\plot 14.097 11.652 15.050 10.351 /
\plot 15.113 10.192 15.621  9.557 /
\plot 16.224 14.224 16.002 13.113 /
\plot 16.002 13.113 16.002 13.208 /
\plot 16.288 15.462 17.177 15.748 /
\plot 16.288 15.462 16.129 16.351 /
\plot 16.066 15.304 15.176 14.732 /
\plot 15.176 14.732 14.034 15.018 /
\plot 14.034 15.018 14.669 15.684 /
\plot 14.669 15.621 16.066 15.304 /
\plot 14.669 15.748 16.066 15.462 /
\plot 16.066 15.526 16.002 16.192 /
\plot 14.034 14.891 15.113 14.573 /
\plot 15.113 14.573 14.541 13.843 /
\plot 14.541 13.843 14.034 14.891 /
\plot 13.875 14.796 14.383 13.780 /
\plot 13.208 13.938 14.319 13.780 /
\plot 13.208 13.780 14.319 13.652 /
\plot 13.303 13.938 13.875 14.796 /
\plot 14.541 15.748 13.875 15.113 /
\plot 13.875 15.113 12.922 15.526 /
\plot 12.922 15.399 13.811 14.954 /
\plot 13.811 14.954 13.144 14.002 /
\plot 13.144 14.002 12.478 14.161 /
\plot 12.478 14.161 12.129 14.732 /
\plot 14.669 15.748 14.605 16.256 /
\plot 14.541 15.748 14.446 16.192 /
\plot 12.478 14.002 13.081 13.843 /
\plot 13.081 13.843 12.700 12.986 /
\plot 12.700 12.986 12.192 13.335 /
\plot 12.192 13.335 12.478 14.002 /
\plot 12.478 14.002 12.478 14.002 /
\plot 14.891 10.287 14.034 11.525 /
\plot 14.034 11.525 13.589 10.192 /
\plot 13.589 10.192 14.541  9.779 /
\plot 14.541  9.779 14.891 10.287 /
\plot 12.351 11.303 13.875 11.589 /
\plot 13.875 11.589 13.430 10.287 /
\plot 13.430 10.287 12.351 11.303 /
\plot 14.954 10.128 15.399  9.557 /
\plot 14.954 10.065 14.669  9.620 /
\plot 13.589 10.065 14.541  9.620 /
\plot 13.589 10.065 14.034  9.557 /
\plot 13.430 10.065 13.811  9.557 /
\plot 13.430 10.065 12.764  9.461 /
\plot 12.700  9.620 13.208 10.065 /
\plot 12.256 11.144 13.303 10.192 /
\plot 11.970 10.065 13.208 10.192 /
\plot 13.208 10.065 12.033  9.906 /
\plot 12.256 11.144 11.970 10.509 /
\plot 12.129 11.239 11.398 11.239 /
\plot 12.129 11.239 11.906 10.636 /
\plot 11.462 12.383 12.129 11.366 /
\plot 12.129 11.366 11.398 11.366 /
\plot 11.525 12.541 12.192 11.589 /
\plot 11.525 13.271 12.033 13.335 /
\plot 12.033 13.430 12.351 14.065 /
\plot 12.351 14.065 11.970 14.669 /
\plot 14.097 11.874 14.446 13.557 /
\plot 13.938 11.874 14.319 13.652 /
\plot 12.859 12.827 13.875 11.874 /
\plot 12.859 12.922 13.208 13.780 /
\plot 14.605 13.716 15.176 14.446 /
\plot 15.272 14.510 16.224 14.224 /
\plot 15.335 14.669 16.224 14.383 /
\plot 15.335 14.669 16.129 15.176 /
\plot 16.129 15.176 16.224 14.383 /
\plot 14.097 11.874 15.176 12.383 /
\plot 15.494 12.319 16.574 11.366 /
\plot 12.351 11.430 12.764 12.700 /
\plot 12.192 11.589 12.636 12.764 /
\plot 12.129 13.208 12.636 12.827 /
\plot 12.764 12.700 13.716 11.811 /
\plot 12.351 11.430 13.875 11.748 /
\plot 12.033 13.208 11.620 13.113 /
\setsolid
\plot 15.494 13.494 15.843 12.256 /
\plot 15.843 12.256 15.843 12.319 /
\plot 15.621 13.557 16.002 12.319 /
\plot 15.843 12.256 15.176 11.366 /
\plot 15.621 13.557 16.796 13.716 /
\plot 15.621 13.716 16.796 13.843 /
\plot 16.796 13.716 16.002 12.319 /
\plot 14.827 14.669 15.494 13.780 /
\plot 14.669 14.573 15.399 13.652 /
\plot 15.494 13.494 14.827 12.541 /
\plot 15.399 13.652 14.764 12.700 /
\plot 16.002 12.160 16.002 12.160 /
\plot 16.002 12.160 15.272 11.239 /
\plot 13.589 13.049 14.764 12.700 /
\plot 13.494 12.922 14.669 12.541 /
\plot 14.827 12.541 15.176 11.366 /
\plot 14.669 12.541 15.050 11.303 /
\plot 13.494 12.922 12.636 12.033 /
\plot 13.367 12.986 12.541 12.160 /
\plot 13.589 14.224 13.589 13.049 /
\plot 13.430 14.287 13.430 13.113 /
\plot 12.636 12.033 12.922 10.859 /
\plot 12.922 10.859 14.161 10.509 /
\plot 14.161 10.509 15.050 11.303 /
\plot 13.589 14.224 14.669 14.573 /
\plot 14.669 14.732 13.589 14.383 /
\plot 13.589 14.383 12.986 14.891 /
\plot 13.430 14.287 12.922 14.732 /
\plot 14.827 14.669 14.954 15.399 /
\plot 14.669 14.732 14.827 15.462 /
\plot 12.986 14.891 13.652 15.684 /
\plot 13.652 15.684 14.827 15.462 /
\plot 15.494 13.780 15.621 14.669 /
\plot 15.621 13.716 15.780 14.669 /
\plot 14.954 15.399 15.621 14.669 /
\plot 15.780 14.669 16.954 14.796 /
\plot 16.796 13.843 16.954 14.796 /
\plot 15.113 11.144 14.319 10.414 /
\plot 13.716 15.748 14.891 15.526 /
\plot 13.716 15.748 15.113 15.843 /
\plot 14.891 15.526 15.113 15.843 /
\plot 15.716 14.796 16.954 14.954 /
\plot 15.113 15.526 15.335 15.843 /
\plot 16.954 14.954 17.177 15.240 /
\plot 15.335 15.843 16.510 15.907 /
\plot 16.510 15.907 17.177 15.304 /
\plot 15.113 15.526 15.716 14.796 /
\plot 13.430 13.113 12.922 13.557 /
\plot 12.922 14.732 12.922 13.557 /
\plot 16.002 12.160 17.304 12.160 /
\plot 17.304 12.160 18.034 11.303 /
\plot 15.272 11.239 15.907 10.287 /
\plot 15.907 10.287 17.304 10.287 /
\plot 17.304 10.287 18.034 11.303 /
\plot 16.859 13.494 17.240 12.319 /
\plot 16.224 12.319 16.859 13.494 /
\plot 11.970 12.478 12.478 12.033 /
\plot 12.478 12.033 12.764 10.859 /
\plot 12.764 10.859 12.541 10.351 /
\plot 12.541 10.351 11.970 10.859 /
\plot 11.970 10.859 11.684 11.970 /
\plot 11.970 12.478 11.684 11.970 /
\plot 12.033 12.605 12.859 13.430 /
\plot 12.859 13.430 13.367 12.986 /
\plot 12.033 12.605 12.541 12.160 /
\plot 15.113 11.144 15.780 10.192 /
\plot 15.780 10.192 14.954  9.461 /
\plot 14.319 10.414 14.954  9.461 /
\plot 12.922 10.700 14.097 10.351 /
\plot 14.097 10.351 14.161  9.906 /
\plot 12.700 10.287 13.208  9.779 /
\plot 13.208  9.779 14.161  9.906 /
\plot 12.922 10.700 12.700 10.287 /
\plot 16.224 12.319 17.240 12.319 /
\plot 12.764 14.732 12.764 13.652 /
\plot 12.764 13.652 11.811 14.065 /
\plot 11.811 14.065 12.764 14.732 /
\plot 11.748 13.938 12.700 13.494 /
\plot 12.700 13.494 11.970 12.764 /
\plot 11.970 12.764 11.748 13.938 /
\plot 11.620 13.938 11.811 12.764 /
\plot 11.811 12.541 11.620 12.097 /
\plot 11.620 12.097 11.303 12.256 /
\plot 11.525 11.970 11.303 12.097 /
\plot 11.525 11.970 11.811 10.859 /
\plot 17.081 14.796 16.954 13.843 /
\plot 11.811 12.700 11.620 12.827 /
\plot 11.811 12.541 11.620 12.700 /
\plot 11.620 13.938 11.398 13.494 /
\plot 11.525 14.065 11.239 13.557 /
\plot 16.954 13.843 17.907 13.780 /
\plot 17.399 12.383 17.971 12.319 /
\plot 17.462 12.256 17.971 12.160 /
\plot 17.462 12.256 18.129 11.430 /
\plot 18.129 11.430 18.320 11.748 /
\plot 18.415 11.589 18.256 11.366 /
\plot 18.256 11.366 18.701 11.303 /
\plot 18.193 11.239 18.701 11.144 /
\plot 18.193 11.239 17.462 10.192 /
\plot 17.462 10.192 17.590  9.842 /
\plot 15.907 10.128 17.304 10.128 /
\plot 17.304 10.128 17.462  9.779 /
\plot 15.907 10.128 15.113  9.398 /
\plot 15.176 15.970 15.272 16.415 /
\plot 15.335 15.970 15.399 16.415 /
\plot 15.335 15.970 16.447 16.034 /
\plot 17.081 14.796 17.304 15.176 /
\plot 17.304 15.176 17.462 15.176 /
\plot 16.637 15.970 17.304 15.399 /
\plot 16.637 15.970 16.796 16.129 /
\plot 16.510 16.034 16.637 16.192 /
\plot 17.399 15.304 17.526 15.304 /
\plot 14.319 10.128 14.827  9.398 /
\plot 14.319 10.128 14.383  9.779 /
\plot 14.383  9.779 14.605  9.461 /
\plot 13.303  9.620 14.161  9.779 /
\plot 14.161  9.779 14.383  9.461 /
\plot 11.906 10.700 12.478 10.192 /
\plot 12.478 10.192 12.351 10.065 /
\plot 12.636 10.128 13.144  9.620 /
\plot 13.144  9.620 13.081  9.461 /
\plot 13.303  9.620 13.208  9.461 /
\plot 12.636 10.128 12.414  9.906 /
\plot 11.811 10.859 11.684 10.795 /
\plot 11.906 10.700 11.748 10.636 /
\plot 12.922 15.018 13.494 15.748 /
\plot 13.494 15.748 13.081 15.907 /
\plot 12.922 15.018 12.541 15.176 /
\plot 12.764 14.891 12.351 15.113 /
\plot 11.684 14.161 12.700 14.891 /
\plot 11.684 14.161 11.525 14.510 /
\plot 11.525 14.065 11.462 14.287 /
\plot 13.811 15.907 15.176 15.970 /
\plot 13.716 15.907 13.811 16.256 /
\plot 13.589 15.907 13.652 16.192 /
\plot 13.589 15.907 13.303 15.970 /
\plot 17.018 13.716 17.399 12.383 /
\plot 17.018 13.716 17.907 13.652 /
\linethickness=1pt
\plot 13.938 11.779  8.541 11.462 /
\linethickness=1pt
\plot 15.526 13.684  9.335 15.272 /
\linethickness=1pt
\plot 12.668 12.891  8.858 13.367 /
\linethickness=1pt
\plot 14.732 14.637 10.128 16.859 /
\put {$t_3$} [lB] at  9.176 16.859
\put {$t_5$} [lB] at  8.541 15.272
\put {$t_4^*$} [lB] at  7.906 13.367
\put {$t_6^*$} [lB] at  7.429 11.303
\linethickness=0pt
\putrectangle corners at  7.429 17.272 and 18.701  9.398
\endpicture
\vskip 20pt
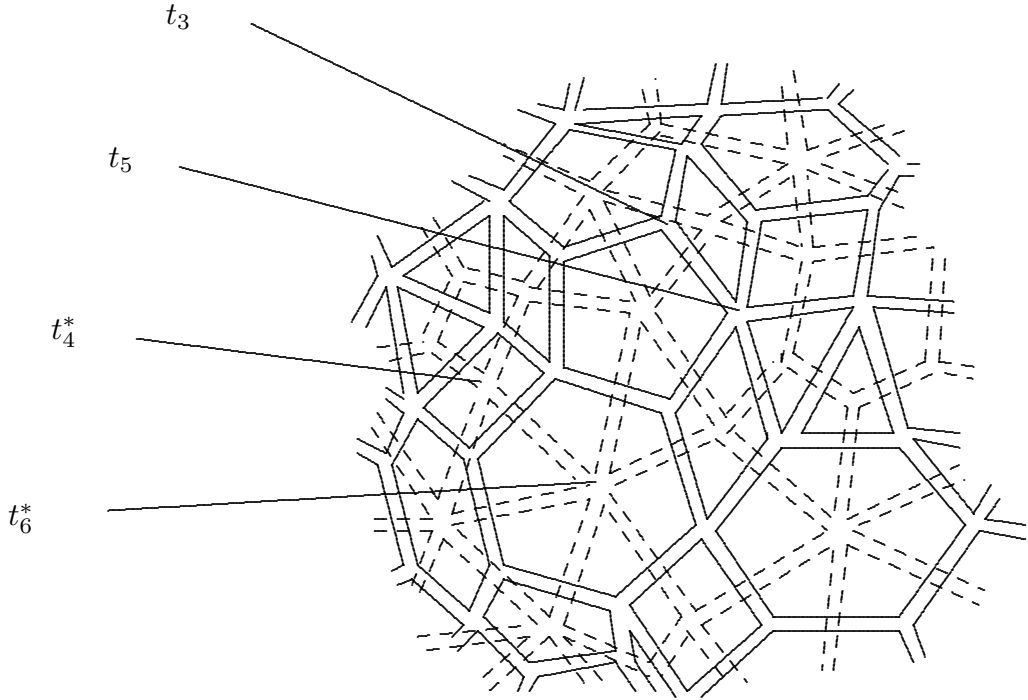
\centerline{{\bf Fig. 1:} A typical surface and some of the associated
weights}
\vskip 20pt
It is clear that this model opens the way to understanding the very
interesting transition mentioned above. If we set $t_4=t_4^*=1$ and
$t_q=t_q^*=0$, for $q\ne 4$, only regular square lattices (graphs)
will exist in equation \DWG . Hence, there are trajectories in the
coupling space of this model, interpolating between pure gravity (for
example, when all $t_q^*=1$) and the regular ``flat'' lattice.

We will show in this paper, that the underlying matrix model
describing the DWG is solvable. Our solution is based on an elegant
representation of this model in terms of the group character expansion
found in \IDiF. It allows us to reduce the $N^2$ degrees of freedom of
the original matrix model to the $N$ degrees of freedom labeling a
representation. We then apply the saddle point approximation to find
the most probable group representation in the corresponding sum over
characters, specified by the distribution of its highest weights.  A
similar approximation was first successfully used in
\DougKaz\  for the calculation of the $QCD_2$ partition function on the
sphere.  We conclude with a well defined (though complicated) integral
equation for this distribution. Though we have not yet been able to
extract the physical picture corresponding to the ``flattening
transition'' we demonstrate on a number of simpler examples that our
method is consistent and correct.

Furthermore, we solve a model apparently inaccessible by standard
methods: we calculate the number of planar graphs having only even
number of neighbours for both original and dual vertices
$t_{2q}=t_{2q}^*=1$, and $ t_{2q-1}=t_{2q-1}^*=0$, for any $q$.

We hope that our methods will lead to new progress in solving many
physically interesting 2D systems. A natural step forward would be the
introduction of matter on DWG, a tempting opportunity, whose success
is, of course, not automatically guarantied. Since our matrix model is
a generalized matrix external field problem, it could also be useful
for new studies in random (mesoscopic) systems.

We present below explicit details of our technique as we feel it is a
general and powerful method for matrix models.

\def\D{\Delta}
\def\sgn{{\rm sgn}}

\newsec{Reduction of the DWG model to a sum over characters}
The partition function for the dually weighted graphs can be
formulated as the following matrix model (see for example \DDSW\ ):
\eqn\DWGmatrix{
Z(t,t^{\ast})=\lambda^{-{N^2\over 2}}\int\,{\cal D}M\ e^{-{N\over 2\lambda} \Tr
M^2\ +\
\sum_{k=1}^{\infty}{1\over k} \Tr B^k\ \Tr (M A)^k}.}
The matrices $A$ and $B$ are external matrices. In the perturbative
expansion of the above integral, the matrix $B$ weights a vertex of
coordination number $q$ with the factor $\Tr\ B^q$ while each face
bounded by $q$ vertices is weighted by a factor $\Tr\ A^q$.
We can therefore make the connection
\eqn\tqAB{
t_q={1\over q}\ {1\over N}\ \Tr\ B^q
{\rm \hskip 20pt and \hskip 20pt}
t_q^{\ast}={1\over q}\ {1\over N}\ \Tr\ A^q.}
Note that it is impossible to solve the above matrix integral by
standard methods since it is unclear, for $A \neq 1$, how to perform
the angular integration, ie. how to evaluate unitary matrix integrals
of the form
$\int\,(d\Omega)_H\exp(\sum_k \beta_k\Tr (M\Omega A\Omega^{\dagger})^k)$
with $M$ and $A$ diagonal.
We circumvent this difficulty by expanding the potential in terms of
the characters on the group:
\eqn\potchar{
e^{\Sigma_{k=1}^{\infty} { 1 \over k} \Tr B^k\ \Tr (MA)^k}=
\prod_{i,j=1}^N{1\over (1-B_i(MA)_j)}
={1\over N^N}\sum_R\chi_R(B)\ \chi_R(MA).}
Here $B_i$ and $(MA)_j$ are the eigenvalues of the matrices $B$ and
$MA$. The first step involves rewriting the sum over $k$ as a double
sum over all the eigenvalues of the matrices $B$ and $MA$ of
$-\ln(1-B_i(MA)_j)$. Exponentiating the log then gives the product
in the numerator. The second step uses
a group theoretic result to write the product in terms
of a sum over characters. The character is defined by the Weyl formula:
\eqn\eigchar{
\chi_{\{h\}}(A)={det_{_{\hskip -2pt (k,l)}}(a_k^{h_l})\over
\Delta(a)},}
where the set of $\{h\}$ are a set of ordered, increasing,
non-negative integers, $\D(a)$ is the Vandermonde determinant, and the
sum over $R$ is the sum over all such sets.  The $R$'s label
representations of the group $U(N)$ and the sets of integers, $\{h\}$,
have the correspondence with the Young tableaux shown in Fig. 2.
\vskip 30pt
\hskip 50 pt

\beginpicture
\setcoordinatesystem units < 1.000cm, 1.000cm>
\linethickness=1pt
\putrectangle corners at  8.414 19.558 and  8.795 19.145
\putrectangle corners at  8.795 19.558 and  9.176 19.145
\putrectangle corners at  9.176 19.558 and  9.557 19.145
\putrectangle corners at  9.557 19.558 and  9.938 19.145
\putrectangle corners at  9.938 19.558 and 10.319 19.145
\putrectangle corners at 10.319 19.558 and 10.700 19.145
\putrectangle corners at 10.700 19.558 and 11.113 19.145
\putrectangle corners at  8.414 19.145 and  8.795 18.764
\putrectangle corners at  8.795 19.145 and  9.176 18.764
\putrectangle corners at  9.176 19.145 and  9.557 18.764
\putrectangle corners at  9.557 19.145 and  9.938 18.764
\putrectangle corners at  8.414 18.764 and  8.795 18.383
\putrectangle corners at  8.795 18.764 and  9.176 18.383
\putrectangle corners at  8.414 18.383 and  8.795 18.034
\putrectangle corners at  8.414 18.034 and  8.795 17.621
\putrule from  8.477 17.621 to  8.414 17.621
\putrule from  8.414 17.621 to  8.414 17.272
\putrule from  8.414 17.272 to  8.477 17.272
\putrule from  8.477 17.272 to  8.414 17.272
\putrule from  8.414 17.272 to  8.414 16.859
\putrule from  8.414 16.859 to  8.477 16.859
\putrule from  8.477 16.859 to  8.414 16.859
\putrule from  8.414 16.859 to  8.414 16.447
\putrule from  8.414 16.447 to  8.477 16.447
\putrule from  8.477 16.447 to  8.414 16.447
\putrule from  8.414 16.447 to  8.414 16.097
\putrule from  8.414 16.097 to  8.477 16.097
\putrule from  8.477 16.097 to  8.414 16.097
\putrule from  8.414 16.097 to  8.414 15.684
\putrule from  8.414 15.684 to  8.477 15.684
\put {$i=1$} [lB] at  7.271 15.684
\put {$i$} [lB] at  7.906 17.653
\put {i  =  N} [lB] at  7.271 19.241
\put {$h_i=i-1+\#$ boxes in row $i$} [lB] at  9.493 17.653
\linethickness=0pt
\putrectangle corners at  7.271 19.558 and 11.113 15.684
\endpicture
\vskip 30pt
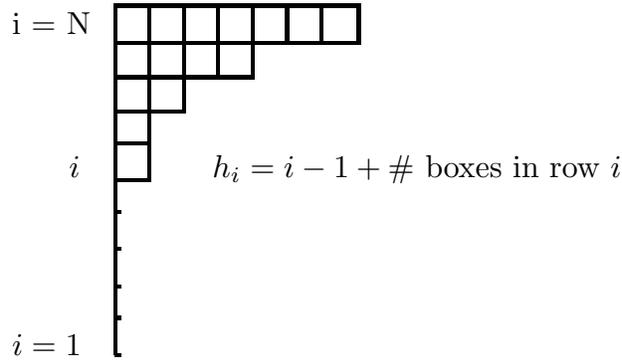
\centerline{{\bf Fig. 2:} Connection between young tableaux and the
integers $h_i$}
\vskip 20pt
\hskip -20pt
Note
that the restriction on the allowed Young tableaux that any row must
have at least as many boxes as the row below implies that the
$\{h_i\}$ are a set of increasing integers
\eqn\hconstr{
h_{i+1}>h_i.}
Substituting
equation \potchar\  into the integral in equation \DWGmatrix\  we can
now do the angular integration using the identity $\int\,({\cal
D}\Omega)_H\chi_R(\Omega M\Omega^{\dagger}A)=
\chi_R(M)\chi_R(A)/d_R$
(where $d_R$ is the dimension of the representation given by
$d_R=\D(h)/\prod_{i=1}^{N-1}i!$), and arrive at the expression
\eqn\ZcharI{
Z={\lambda^{{-N^2\over 2}}\over N^N}\sum_R{1\over d_R}\chi_R(A)\ \chi_R(B)
\int\,\prod_{i=1}^N {\cal D}M_i\,\,\D(M)det_{_{\hskip -2pt (k,l)}}(M_k^{h_l})
e^{-{N\over 2\lambda} \Tr M^2}.}
The gaussian integral can be done explicitly and we arrive at the
final formula
\eqn\IzDiFr{
Z=c\,\sum_{\{h^e,h^o\}}
{\prod_i(h^e_i-1)!!h^o_i!!\over
\prod_{i,j}(h^e_i-h^o_j)}\chi_{\{h\}}(A)\chi_{\{h\}}(B)
\bigl({\lambda\over N}\bigr)^{-{1\over 4}N(N-1)+{1\over
2}\sum_i(h^e_i+h^o_i)},}
where $c$ is some numerical constant that we can drop, the
$\{h^e\}$ are a set of $N/2$ even, increasing, non-negative
integers and the $\{h^o\}$ are $N/2$ odd, increasing, positive
integers, and the sum is over all such sets. In other words the
original sum is now restricted to the subsets of $\{h\}$ with equal
numbers of even and odd integers. This is an exact result. The sum is
in general divergent, as is the original matrix integral, and should
be thought of as a generating function for graphs of arbitrary genus.
This formula was originally derived by Itzykson and Di Francesco \IDiF\
by summing up ``fatgraphs''. For the rest of this paper we restrict our
attention solely to the genus 0 contribution. In other words we will
be studying the large $N$ limit of equation \IzDiFr .

There is a second useful formula for the character given in terms of
Schur polynomials, $P_n(\theta)$, defined by
\eqn\schurpolyn{
e^{\Sigma_{i=1}^{\infty} z^i\theta_i}=\sum_{n=0}^{\infty}z^n
P_n(\theta).}
In terms of Schur polynomials the character is
\eqn\schurchar{
\chi_{\{h\}}(A)=det_{_{\hskip -2pt
(k,l)}}\bigl(P_{h_k+1-l}(\theta)\bigr),}
where $\theta_i={1\over i} \Tr A^i$.  In general, the explicit
expressions for the characters are very complicated. Certain specific
cases however yield simple results which we will state as we need
them.

\newsec{Relations between highest weight and matrix model quantities}
Before we look for the large $N$ limit of equation \IzDiFr , we derive
some explicit expressions relating useful quantities from matrix
models to quantities encountered in the language of highest weights. In
the large $N$ limit of equation \IzDiFr , we assume that the sum over
all representations will be dominated by a single contribution, or a
single Young tableau, $\{h_i\}$, and introduce a density $\rho(h)$
defined in the standard way by $\rho(h)={1\over N}{\partial
i\over\partial h}$.  We are being slightly sloppy with the notation
here since to define a sensible density we have to rescale the
integers $\{h_i\}$ by dividing them by $N$. For the rest of this paper
an $h_i$ with an index refers to one of the original integers and $h$
without a subscript refers to the rescaled continuous parameter.

All the formulae in this section have their root in the simple
observation that
\eqn\trcharobs{
\Tr A^q=\sum_k{\chi_{\{\tilde h\}}(A)\over\chi_{\{h\}}(A)}
{\rm \hskip 20pt where \hskip 20pt}
\tilde h_i=h_i+q\delta_{i,k}.}
For compactness of notation we have omitted labeling the $\tilde h$
with an index $k$.  This formula follows directly from the Weyl
formula for the character, \eigchar. The character can be written in
terms of the Itzykson Zuber integral,
$I(h,\alpha)= det_{_{\hskip -2pt
(k,l)}}(e^{h_k\alpha_l})/(\D(h)\D(\alpha))$, as
\eqn\charIZ{
\chi_{h}(A)=I(h,\alpha)\D(h){\D(\alpha)\over\D(a)},}
where $\alpha_l$ is defined through the eigenvalues of $A$ by
$a_l=e^{\alpha_l}$. This allows us to write
\eqn\trcharsteps{
\Tr A^q=\sum_k{\D(\tilde h)\over\D(h)}
e^{q{\ln I(\tilde h,\alpha)-\ln I(h,\alpha)\over q}}
\sim\lim_{N\rightarrow\infty}
\sum_k\prod_{j(\neq k)}\bigl(1+{q\over h_k-h_j}\bigr)
e^{q F(h_k)},}
where
\eqn\IZdef{F(h_k)={\partial\over\partial h_k}log I(h,\alpha)}
is the derivative of the Itzykson Zuber integral. In the last step we
assume that the Itzykson Zuber integral has a well defined large $N$
limit: $I(h,\alpha)=e^{N^2 F_0[\rho(h),\rho(\alpha)] + O(N^0)}$.  We
now notice that we can replace the sum by a contour integral in $h$
(encircling all the $h_k$'s) if we also unrestrict the product
allowing it to be a product over all $j$. This contour integration
trick was originally invented in \VKaz\ for a much simpler model
($c=-2$ gravity), and more recently was used in \KW\ for the large $N$
limit of the heat kernel. Applying it in this case we obtain:
\eqn\trchar{
{1\over N} \Tr A^q={1\over q}\oint\,{dh\over 2\pi i}
e^{q(H(h)+F(h))}
{\rm \hskip 20pt with \hskip 20pt}
H(h)=\int dh'\ {\rho(h') \over h-h'}.}
$F(h)$ is initially defined only on the support of $h$; we have then
analytically continued $F(h)$ into the whole complex plane, so that
the contour integral, which circles the cut of the resolvent, $H(h)$,
is well defined. We have thus rederived in a very compact way the
large $N$ limit of the Itzykson Zuber integral \Mig\ \Matyt\ . To make
the connection with the result in \Mig\ more explicit it is simple to
expand \trchar\ as a power series in $q$ and then resum the series to
express the result in terms of the resolvent for the eigenvalues
$\alpha$ (see also \KW\ ):
\eqn\Mig{
\Theta(\alpha)=-\oint\,{dh\over 2\pi i}\ln(\alpha-H(h)-F(h))
{\rm \hskip 20pt with \hskip 20pt}
\Theta(\alpha)=\int\,{d\alpha'\sigma(\alpha')\over\alpha-\alpha'},}
where $\sigma(\alpha)$ is the density for the eigenvalues $\alpha$.

Next we look at the expectation value of ${1\over N} \Tr M^{2q}$.
Placing this term into the $M$ integrand in equation \ZcharI\ and
using similar steps as for ${1\over N} \tr A^q$ we arrive at
\eqn\trMtwoq{
{1\over N} \Tr M^{2q}={\lambda^q\over q}\oint\,{dh\over 2\pi i}h^qe^{qH(h)}.}
This formula is derived only for traces of even powers of the matrix
$M$. The expectation value of ${1\over N} \Tr (MA)^{2q}$ is derived in
a like manner. This time we substitute ${1\over N} \Tr (MA)^{2q}$ into
an earlier step in the derivation of the Itzykson-Di Francesco
formula, namely into equation \potchar . Using \trcharobs\ we see that
this time we shift both $\chi_R(A)$ and $\chi_R(M)$, which leads to
the final result
\eqn\trMAtwoq{
{1\over N}\Tr ((MA)^{2q})={\lambda^q\over q}\oint\,{dh\over 2\pi i}h^q
e^{q(H(h)+2F(h))}.}
Again this is derived only for traces of the even powers of $MA$. We
assume from here on that we are working with an even potential so that
it is only the even traces that remain.  Summing up equations
\trMtwoq\ and \trMAtwoq\ over all $q$, assuming also that the
potential is even, we arrive at two formulae for the two types of
resolvents one can define for the original matrix model:
\eqn\resolvents{\eqalign{
W(P)=&\langle {1\over N} \Tr {1\over P-M}\rangle ={1\over P}-
                                     {1\over P}\oint\,{dh\over 2\pi i}
\ln(P^2-\lambda he^{H(h)})\cr
W_A(P)=&\langle {1\over N} \Tr {1\over P-MA}\rangle ={1\over P}-
                                     {1\over P}\oint\,{dh\over 2\pi i}
\ln(P^2-\lambda he^{H(h)+2F(h)})}.}
The first equation can be solved by Lagrange inversion since we know
that the only singularity of $H(h)$ is the cut circled by the
contour. Performing the inversion gives the very simple pair of
equations
\eqn\resolvinvers{\eqalign{
PW(P)=&{P^2\over\lambda}-h\cr
\lambda he^{H(h)}=&P^2.}}
Knowing $H(h)$ we perform a functional inversion to obtain the
resolvent of the original matrix model. We cannot in general do the
same inversion for the resolvent $W_A(P)$ since we do not know in
advance the singularities of $F(h)$.

\newsec{The gaussian model: straightening random loops}

We will now check the power of our method on the simplest non-trivial
case of our external field problem: We simply set $B=0$, i.e.  all
$t_q=0$ in equation \DWGmatrix. Now there is no potential at all and
thus no weights are excited: The only contribution to equation
\IzDiFr\ is the empty Young tableau:
\eqn\flat{H(h)=\ln {h \over h-1}.}
One immediately checks that the inversion formulae \resolvinvers\
correctly give the Wigner semi-circle law (we may set $\lambda=1$ here)
\eqn\wigner{W(P)={1 \over 2} \big( P - \sqrt{P^2-4} \big).}
Clearly $W(P)$ and $H(h)$ are unchanged even in the presence of
non-trivial coupling constants $t_q^*$, but now the interesting
quantity is the resolvent $W_A(P)$:
\eqn\wa{W_A(P)={1  \over Z}\ \int\,{\cal D}M\ e^{-{N\over 2} \Tr M^2}\
{1\over N} \Tr {1\over P-MA}.}
We can find it by substituting $H(h)$ into \trchar
\eqn\mig{\sum_{q=1}^{\infty}\ q t_q\ \omega^q=-\oint\,{dh\over 2\pi i}\
\ln \big[ h-1-\omega\ h\ e^{F(h)} \big] -1.}
where we also summed up the moments constructing their generating
function with an auxiliary variable $\omega$. One now sees that in
this case the Lagrange inversion can be performed by picking up
a pole term {\it inside} the contour, giving immediately
\eqn\miglag{h\ -1=\sum_{q=1}^{\infty}\ q t_q^*\ \omega^q
{\rm \hskip 20pt and \hskip 20pt}
h\ -1= \omega\ h\ e^{F(h)}.}
In addition, from the inversion formula \resolvents\ for $W_A(P)$ we find,
using the same method,
\eqn\reslag{h=P\ W_A(P)
{\rm \hskip 20pt and \hskip 20pt}
h\ -1= {1 \over P^2}\ h^2\ e^{2\ F(h)}.}
Eliminating $F(h)$ we find the exact solution of our problem:
\eqn\sol{P^2\ \omega^2\ =\sum_{q=1}^{\infty}\ q t_q^*\ \omega^q
{\rm \hskip 20pt and \hskip 20pt}
P\ W_A(P) = 1\ + P^2\ \omega^2.}
Indeed, this set of functional equations determines, for any set of
couplings $t_q^*$, after elimination of $\omega$, the desired
resolvent.  Let us remark that these equations can alternatively be
derived using Schwinger-Dyson techniques, yielding a non-trivial check
of our functional methods. It is interesting to observe that we may
obtain arbitrarily complicated resolvents by freely choosing the
$t_q^*$'s. On the other hand, it is seen that a finite number of
non-zero coupling constants always leads to an algebraic resolvent.
An amusing toy system consists in only activating the first three
coupling constants. Now the resolvent $W_A(P)$ is interpreted as the
generating function of rainbow graphs with face-valency not larger
than three, see Fig. 3 below.  It is, from \sol, given explicitly by
\eqn\rain{W_A(P)=
{1 \over P} + {P \over 2 {t_3^*}^2}\ \bigg[
(P^2 - t_2^*)^2 - 2 t_1^* t_3^* -
\sqrt{(P^2 - t_2^*)^2 - 4 t_1^* t_3^*} \bigg] .}
It is interesting to investigate what happens in this toy system if we
tune away the faces with negative boundary curvature, i.e.
$t_3 \rightarrow 0$:
\eqn\cigar{W_A(P)={1 \over P}
+ {P\ {t_1^*}^2 \over (P^2 - t_2^*)^2}.}
Now we obtain merely the ``cigar-like'' diagrams below.
\vskip 20pt
\beginpicture
\setcoordinatesystem units < 1.000cm, 1.000cm>
\linethickness=1pt
\circulararc 186.734 degrees from 17.526 18.320 center at 17.311 18.460
\circulararc 88.969 degrees from 17.050 18.510 center at 17.401 18.642
\circulararc 72.314 degrees from 16.923 18.224 center at 17.303 18.380
\circulararc 58.109 degrees from 16.732 17.907 center at 17.167 18.151
\circulararc 75.750 degrees from 16.542 17.590 center at 16.907 17.764
\circulararc 61.926 degrees from 16.351 17.240 center at 16.785 17.484
\circulararc 61.926 degrees from 16.161 16.859 center at 16.595 17.103
\circulararc 72.314 degrees from 16.002 16.510 center at 16.382 16.665
\circulararc 58.109 degrees from 15.811 16.192 center at 16.246 16.437
\circulararc 75.750 degrees from 15.621 15.811 center at 15.986 15.986
\circulararc 61.926 degrees from 15.431 15.526 center at 15.864 15.769
\circulararc 180.000 degrees from 14.510 13.811 center at 14.732 13.652
\circulararc 51.539 degrees from 14.573 13.906 center at 15.026 14.256
\circulararc 58.109 degrees from 14.764 14.224 center at 15.199 14.468
\circulararc 72.314 degrees from 14.954 14.541 center at 15.334 14.697
\circulararc 86.143 degrees from 15.081 14.891 center at 15.406 15.000
\circulararc 61.926 degrees from 15.240 15.208 center at 15.674 15.452
\circulararc 185.450 degrees from  8.509 14.510 center at  8.726 14.388
\circulararc 73.740 degrees from  9.081 17.272 center at  8.954 17.367
\circulararc 84.547 degrees from  9.081 17.526 center at  9.163 17.897
\circulararc 77.982 degrees from  9.779 17.621 center at 10.112 17.351
\circulararc 73.740 degrees from 10.763 17.462 center at 10.890 17.367
\circulararc 187.470 degrees from 10.986 17.907 center at 10.772 18.049
\circulararc 64.942 degrees from 10.509 18.066 center at 10.906 18.304
\circulararc 84.547 degrees from 10.319 17.780 center at 10.681 17.897
\circulararc 77.982 degrees from 10.033 17.081 center at  9.700 17.351
\circulararc 70.723 degrees from  9.525 16.923 center at  9.101 16.752
\circulararc 188.800 degrees from  9.430 17.907 center at  9.182 17.798
\circulararc 73.740 degrees from  9.017 16.034 center at  9.144 15.939
\circulararc 54.555 degrees from  8.731 16.637 center at  9.185 16.988
\circulararc 70.723 degrees from  8.572 16.351 center at  8.996 16.522
\circulararc 83.267 degrees from  8.287 15.653 center at  8.001 15.907
\circulararc 73.740 degrees from  7.303 15.843 center at  7.176 15.939
\circulararc 197.945 degrees from  7.525 16.701 center at  7.282 16.609
\circulararc 54.555 degrees from  7.112 16.415 center at  7.135 16.988
\circulararc 70.723 degrees from  7.303 16.066 center at  7.323 16.522
\circulararc 107.415 degrees from  8.033 16.161 center at  8.219 15.907
\circulararc 97.502 degrees from  9.049 15.748 center at  8.922 15.412
\circulararc 182.440 degrees from 10.541 16.542 center at 10.766 16.680
\circulararc 57.684 degrees from 10.954 16.923 center at 11.020 16.360
\circulararc 70.723 degrees from 10.763 17.209 center at 10.742 16.752
\circulararc 53.130 degrees from  9.398 14.796 center at  9.414 14.589
\circulararc 66.906 degrees from  8.985 14.319 center at  8.563 14.150
\circulararc 75.332 degrees from  9.176 14.669 center at  8.789 14.508
\circulararc 84.547 degrees from  8.731 15.240 center at  8.814 15.611
\circulararc 180.000 degrees from  9.906 14.827 center at  9.906 15.081
\circulararc 68.762 degrees from  9.842 15.335 center at 10.198 15.058
\circulararc 82.222 degrees from  9.462 15.335 center at  9.753 15.081
\circulararc 189.148 degrees from  7.080 15.399 center at  7.291 15.238
\circulararc 70.723 degrees from  7.557 15.176 center at  7.133 15.006
\circulararc 97.502 degrees from  7.747 15.494 center at  7.397 15.412
\circulararc 57.684 degrees from  9.335 16.605 center at  8.824 16.360
\plot 17.081 18.574 14.510 13.811 /
\plot 17.526 18.320 14.954 13.494 /
\plot 11.874 16.066 14.256 16.066 /
\plot 13.938 16.224 14.256 16.066 /
\plot 14.256 16.066 13.938 15.907 /
\plot  7.303 16.669  6.445 17.653 /
\plot  7.493 16.320  7.493 16.320 /
\plot  7.493 16.320  6.001 16.542 /
\plot  7.588 15.939  6.001 15.526 /
\plot  9.207 14.700  8.954 14.287 /
\plot  9.557 17.716  9.620 17.653 /
\plot  9.620 17.653  9.716 17.621 /
\plot  9.557 16.986  9.620 17.050 /
\plot  9.620 17.050  9.716 17.081 /
\plot 10.128 17.621  9.684 17.621 /
\plot 10.287 17.716 10.224 17.653 /
\plot 10.224 17.653 10.128 17.621 /
\plot 10.287 16.986 10.224 17.050 /
\plot 10.224 17.050 10.128 17.081 /
\plot 10.541 18.161 10.287 17.716 /
\plot 10.986 17.907 10.731 17.462 /
\plot  8.858 16.828  9.081 17.240 /
\plot  9.303 16.542  9.557 16.986 /
\plot  8.985 17.653  9.081 17.462 /
\plot  9.430 17.939  9.557 17.716 /
\plot  9.716 17.081 10.128 17.081 /
\plot  8.350 16.161  7.906 16.161 /
\plot  8.541 16.288  8.477 16.224 /
\plot  8.477 16.224  8.350 16.161 /
\plot  8.541 15.558  8.477 15.621 /
\plot  8.477 15.621  8.350 15.653 /
\plot  8.795 16.701  8.541 16.288 /
\plot  9.239 16.447  8.985 16.034 /
\plot  7.938 16.161  8.382 16.161 /
\plot  7.747 16.288  7.811 16.224 /
\plot  7.811 16.224  7.938 16.161 /
\plot  7.747 15.558  7.811 15.621 /
\plot  7.811 15.621  7.938 15.653 /
\plot  7.525 16.701  7.779 16.288 /
\plot  7.080 16.447  7.303 16.034 /
\plot  8.795 15.113  8.541 15.558 /
\plot 10.986 16.828 10.763 17.240 /
\plot 10.541 16.542 10.287 16.986 /
\plot  8.763 14.986  8.795 15.081 /
\plot  8.795 15.081  8.763 15.176 /
\plot  9.398 15.335  9.303 15.367 /
\plot  9.303 15.367  9.239 15.431 /
\plot  8.509 14.541  8.763 14.986 /
\plot  8.763 15.176  8.604 15.462 /
\plot  9.207 15.462  9.081 15.684 /
\plot  9.906 14.827  9.398 14.827 /
\plot  9.906 15.335  9.398 15.335 /
\plot  7.080 15.399  7.303 15.811 /
\plot  7.525 15.113  7.747 15.558 /
\plot  8.890 16.891  8.731 16.605 /
\plot  9.335 16.669  9.239 16.447 /
\plot  7.938 15.653  8.382 15.653 /
\plot  9.017 15.843  9.081 15.684 /
\put {$t_1^*$} [lB] at  5.889 17.812
\put {$t_2^*$} [lB] at  5.413 16.542
\put {$t_3^*$} [lB] at  5.413 15.272
\put {$t_3^*\neq0$} [lB] at  7.429 12.573
\put {$t_3^*=0$} [lB] at 14.891 12.573
\linethickness=0pt
\putrectangle corners at  4.413 18.764 and 17.590 12.573
\endpicture
\vskip 20pt
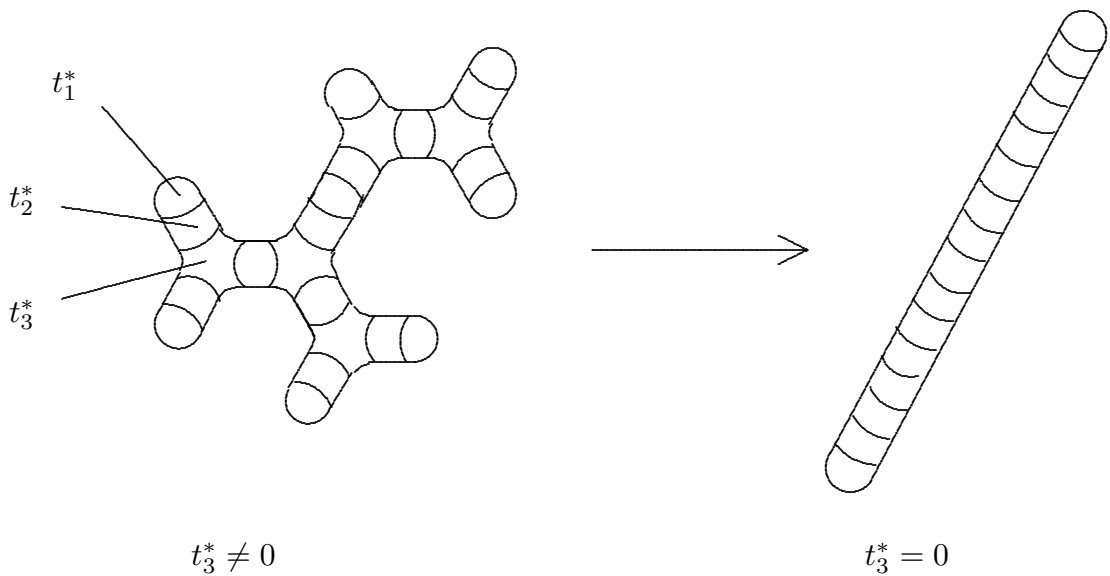
\centerline{{\bf Fig. 3:} ``Rainbow'' $\longrightarrow$ ``cigar-like'' diagrams
in the gaussian model}
\vskip 20pt
In fact, as it was argued in Appendix A of \mss, the universal
continuum limit (with string susceptibility $\gamma_{{\rm
str}}={1\over 2}$
due to the square-root singularity in equation \rain ) of the model
\rain\ can be interpreted as two-dimensional {\it topological} quantum
gravity: The expectation value of the metric tensor is zero in the
bulk, leading to a theory of quantum loops.  ``Flattening'' in such a
theory is thus just ``straightening'', and indeed the continuum limit
of \cigar, Fig. 2, is simply a straightened loop with two curvature
defects. Note that the cross-over from the ``quantum'' to the
``straight'' phase is simply a catastrophe in the algebraic sense: As
soon as we turn on the negative curvature coupling $t_3^*$, the defects
proliferate and the straight line disorders. It is not excluded that a
similar rather trivial mechanism will govern the crossover from random
to flat graphs. However, this is not the most likely scenario; indeed
one is reminded of the fundamental difference between one and
two-dimensional systems with regard to the absence, respectively
presence, of phase transitions. At any rate, let us turn to real planar
graphs.

\newsec{Saddlepoint equations and planar graphs}

Success was guaranteed in the case of the Gaussian model since
we knew from the start the trivial, linear distribution of weights.
We now have to establish that non-trivial distributions $H(h)$ can
be found and that we are able to reproduce planar graphs from
the character expansion. One finds that saddlepoint techniques may
be successfully applied if certain precautions, to be elaborated below,
are taken. Let us illustrate the method and its subtleties on a number
of examples:

\def\J {{\cal J}}
In our new language, the simplest model generating planar graphs turns
out to be the case $A=1$ and $B=\J$, where $\Tr\ \J^q$ equals one if
$q$ is even, and zero otherwise. Thus we obtain a traditional one matrix
model with the ``even-log'' potential $-{1 \over 2} \ln\ (1-M^2)$,
generating planar graphs with arbitrary even vertices. It is easy to
explicitly work out -- with the help of the Schur character formula
\schurchar\  (see Appendix) -- the characters for this case:
\eqn\chars{\chi_{\{h\}}(1)\sim \Delta(h)
{\rm \hskip 20pt and \hskip 20pt}
\chi_{\{h\}}(\J)\sim \Delta(h^o)\ \Delta(h^e)\
\sgn \prod_{i,j}(h_i^e-h_j^o).}
 From here on we will omit irrelevant numerical constants.
One now sees that the character expansion \IzDiFr\ for the partition
function becomes
\eqn\evlog{Z \sim
\lambda^{-{1\over 4}N(N-1)}\sum_{\{h^e,h^o\}}
\prod_i(h^e_i-1)!!h^o_i!!\
\Delta(h^o)^2\ \Delta(h^e)^2\
\bigl({\lambda\over N}\bigr)^
{{1\over 2}\sum_i(h^e_i+h^o_i)}.}
We thus observe that even and odd weights completely factorize! By
symmetry, they should have the same statistical distribution. This
partition sum is ideally suited for a saddlepoint analysis: the
Vandermondes repel the weights from each other while the potential
attracts -- for small coupling $\lambda$ -- to the origin. It is
therefore natural to write down, in the large $N$ limit, the
saddlepoint equation
\eqn\sadevlog{\barint\ dh'\ {\rho(h') \over h-h'} =
-{1 \over 2}\ \ln\ ( h\ \lambda).}
obtained in the standard fashion from equation \evlog. The density
$\rho(h)$ and the continuous variables $h$ were defined in section 3
and one also uses Sterling's formula: $\ln\ h!! \sim {h \over
2}(\ln(h)-1)$.  This equation is easily solved but leads to the {\it
wrong} result. The phenomenon is identical to the one previously
encountered in \DougKaz: The naive saddle point equation fails to take
into account the constraint $\rho(h)\leq 1$ which follows from
equation \hconstr . Imposing the condition that the density is
saturated at its maximum value $\rho(h)=1$ on the interval $[0,b]$, we
write down the modified saddlepoint equation
\eqn\msadevlog{\barint_b^a\ dh'\ {\rho(h') \over h-h'} =
-{1 \over 2}\ \ln\ ( h\ \lambda) - \ln ({h\over h-b})}
determining the non-trivial piece of the density on
the interval $[b,a]$. We generate the
full analytic function $H(h)=\int\ dh'\ \rho(h')/(h-h')$ from
$-\hskip -9pt\int\ dh'\ \rho(h') /(h-h')$ by performing the contour
integral
\eqn\contH{H(h)=
\ln({h\over h-b})+\sqrt{(h-a)(h-b)}\oint_C\ {ds\over 2\pi i}
{{1\over 2}\ln(s\lambda)+\ln ({s\over s-b})\over
(s-h)\sqrt{(s-a)(s-b)}},}
where the contour encircles the cut $[a,b]$. Inflating the contour and
catching instead the cuts $[\infty,0]$ and $[0,b]$ we arrive at
\eqn\Hevlog{H(h)=
\ln\left[{\sqrt{a}-\sqrt{b}\over\sqrt{\lambda}}
{h+\sqrt{ab}+\sqrt{(h-a)(h-b)}\over
(a+b)h-2ab+2\sqrt{ab}\sqrt{(h-a)(h-b)}}\right],}
with
\eqn\abevlog{\eqalign{
\Biggl({\sqrt{a}+\sqrt{b}\over 2}\Biggr)^2=&
{1\over 4\lambda}(1-\sqrt{1-8\lambda})\cr
\Biggl({\sqrt{a}-\sqrt{b}\over 2}\Biggr)^2=&
{1\over 8\lambda}(1-4\lambda-\sqrt{1-8\lambda})},}
the constants $a$ and $b$ being fixed by the condition that
$H(h)=1/h+O(1/h^2)$. This is the solution to the ``even-log'' matrix
model in the language of highest weights, reproducing the correct
critical coupling $\lambda_c={1 \over 8}$.  Below we show a plot of
$\rho(h)$ for various values of $\lambda$.
\vskip 20pt
\beginpicture
\setcoordinatesystem units < 1.000cm, 1.000cm >
\linethickness=1pt
\putrule from 3.8 14 to 13.4 14
\putrule from 4 13.8 to 4 19.2
\putrule from 6. 14 to 6. 14.1
\putrule from 4.4 14 to 4.4 14.05
\putrule from 4.8 14 to 4.8 14.05
\putrule from 5.2 14 to 5.2 14.05
\putrule from 5.6 14 to 5.6 14.05
\putrule from 8. 14 to 8. 14.1
\putrule from 6.4 14 to 6.4 14.05
\putrule from 6.8 14 to 6.8 14.05
\putrule from 7.2 14 to 7.2 14.05
\putrule from 7.6 14 to 7.6 14.05
\putrule from 10. 14 to 10. 14.1
\putrule from 8.4 14 to 8.4 14.05
\putrule from 8.8 14 to 8.8 14.05
\putrule from 9.2 14 to 9.2 14.05
\putrule from 9.6 14 to 9.6 14.05
\putrule from 12. 14 to 12. 14.1
\putrule from 10.4 14 to 10.4 14.05
\putrule from 10.8 14 to 10.8 14.05
\putrule from 11.2 14 to 11.2 14.05
\putrule from 11.6 14 to 11.6 14.05
\putrule from 12.4 14 to 12.4 14.05
\putrule from 12.8 14 to 12.8 14.05
\putrule from 13.2 14 to 13.2 14.05
\putrule from 4 15. to 4.1 15.
\putrule from 4 14.2 to 4.05 14.2
\putrule from 4 14.4 to 4.05 14.4
\putrule from 4 14.6 to 4.05 14.6
\putrule from 4 14.8 to 4.05 14.8
\putrule from 4 16. to 4.1 16.
\putrule from 4 15.2 to 4.05 15.2
\putrule from 4 15.4 to 4.05 15.4
\putrule from 4 15.6 to 4.05 15.6
\putrule from 4 15.8 to 4.05 15.8
\putrule from 4 17. to 4.1 17.
\putrule from 4 16.2 to 4.05 16.2
\putrule from 4 16.4 to 4.05 16.4
\putrule from 4 16.6 to 4.05 16.6
\putrule from 4 16.8 to 4.05 16.8
\putrule from 4 18. to 4.1 18.
\putrule from 4 17.2 to 4.05 17.2
\putrule from 4 17.4 to 4.05 17.4
\putrule from 4 17.6 to 4.05 17.6
\putrule from 4 17.8 to 4.05 17.8
\putrule from 4 19. to 4.1 19.
\putrule from 4 18.2 to 4.05 18.2
\putrule from 4 18.4 to 4.05 18.4
\putrule from 4 18.6 to 4.05 18.6
\putrule from 4 18.8 to 4.05 18.8
\put {1} [1B] at 5.95 13.7
\put {2} [1B] at 7.95 13.7
\put {3} [1B] at 9.95 13.7
\put {4} [1B] at 11.95 13.7
\put {0.2} [1B] at 3.6 14.95
\put {0.4} [1B] at 3.6 15.95
\put {0.6} [1B] at 3.6 16.95
\put {0.8} [1B] at 3.6 17.95
\put {1.} [1B] at 3.6 18.95
\put {$h$} [1B] at 12.55 13.5
\put {$\rho(h)$} [1B] at 3.5 18.6
\plot 5.8997 18. 7.2 18.5 /
\plot 5.7546 17.5 7.2 18 /
\plot 5.5026 17. 7.2 17.5 /
\plot 5.4033 16.5 7.2 17 /
\put {$\lambda= 0.001$} [1B] at 8.1 18.5
\put {$\lambda= 0.01$} [1B] at 8 18.
\put {$\lambda= 0.0625$} [1B] at 8.13 17.5
\put {$\lambda= 0.125$} [1B] at 8.03 17.
\plot 4 19 5.87915 19 /
\plot 5.87915 19 5.88039 18.7558 /
\plot 5.88039 18.7558 5.88397 18.5173 /
\plot 5.88397 18.5173 5.88971 18.2838 /
\plot 5.88971 18.2838 5.89742 18.055 /
\plot 5.89742 18.055 5.90691 17.8304 /
\plot 5.90691 17.8304 5.91797 17.6095 /
\plot 5.91797 17.6095 5.93044 17.392 /
\plot 5.93044 17.392 5.94411 17.1774 /
\plot 5.94411 17.1774 5.9588 16.9652 /
\plot 5.9588 16.9652 5.97431 16.7547 /
\plot 5.97431 16.7547 5.99045 16.5453 /
\plot 5.99045 16.5453 6.00704 16.3362 /
\plot 6.00704 16.3362 6.02388 16.1263 /
\plot 6.02388 16.1263 6.04079 15.9142 /
\plot 6.04079 15.9142 6.05757 15.6979 /
\plot 6.05757 15.6979 6.07403 15.4742 /
\plot 6.07403 15.4742 6.08998 15.2381 /
\plot 6.08998 15.2381 6.10524 14.9791 /
\plot 6.10524 14.9791 6.11961 14.6695 /
\plot 6.11961 14.6695 6.1329 14.0002 /
\plot 5.64996 19 5.65398 18.7029 /
\plot 5.65398 18.7029 5.66563 18.414 /
\plot 5.66563 18.414 5.6843 18.1337 /
\plot 5.6843 18.1337 5.70937 17.8626 /
\plot 5.70937 17.8626 5.74021 17.6007 /
\plot 5.74021 17.6007 5.7762 17.348 /
\plot 5.7762 17.348 5.81674 17.1045 /
\plot 5.81674 17.1045 5.86119 16.8697 /
\plot 5.86119 16.8697 5.90895 16.6433 /
\plot 5.90895 16.6433 5.95938 16.4246 /
\plot 5.95938 16.4246 6.01188 16.2127 /
\plot 6.01188 16.2127 6.06583 16.0069 /
\plot 6.06583 16.0069 6.1206 15.8058 /
\plot 6.1206 15.8058 6.17557 15.6081 /
\plot 6.17557 15.6081 6.23013 15.4119 /
\plot 6.23013 15.4119 6.28366 15.2142 /
\plot 6.28366 15.2142 6.33555 15.0106 /
\plot 6.33555 15.0106 6.38516 14.7927 /
\plot 6.38516 14.7927 6.43188 14.538 /
\plot 6.43188 14.538 6.4751 14. /
\plot 5.24662 19 5.25898 18.5319 /
\plot 5.25898 18.5319 5.29481 18.0845 /
\plot 5.29481 18.0845 5.35219 17.6643 /
\plot 5.35219 17.6643 5.42922 17.2754 /
\plot 5.42922 17.2754 5.52401 16.9196 /
\plot 5.52401 16.9196 5.63466 16.5967 /
\plot 5.63466 16.5967 5.75925 16.3052 /
\plot 5.75925 16.3052 5.89589 16.0428 /
\plot 5.89589 16.0428 6.04267 15.8067 /
\plot 6.04267 15.8067 6.19769 15.5941 /
\plot 6.19769 15.5941 6.35906 15.4022 /
\plot 6.35906 15.4022 6.52486 15.2281 /
\plot 6.52486 15.2281 6.6932 15.0693 /
\plot 6.6932 15.0693 6.86218 14.9233 /
\plot 6.86218 14.9233 7.02988 14.7876 /
\plot 7.02988 14.7876 7.19442 14.6594 /
\plot 7.19442 14.6594 7.35388 14.5355 /
\plot 7.35388 14.5355 7.50637 14.4106 /
\plot 7.50637 14.4106 7.64999 14.273 /
\plot 7.64999 14.273 7.78282 14. /
\plot 5. 19 5.039 18.123 /
\plot 5.039 18.123 5.152 17.3317 /
\plot 5.152 17.3317 5.333 16.6591 /
\plot 5.333 16.6591 5.576 16.1088 /
\plot 5.576 16.1088 5.875 15.667 /
\plot 5.875 15.667 6.224 15.3152 /
\plot 6.224 15.3152 6.617 15.0351 /
\plot 6.617 15.0351 7.048 14.8116 /
\plot 7.048 14.8116 7.511 14.6328 /
\plot 7.511 14.6328 8. 14.4892 /
\plot 8. 14.4892 8.509 14.3739 /
\plot 8.509 14.3739 9.032 14.2812 /
\plot 9.032 14.2812 9.563 14.207 /
\plot 9.563 14.207 10.096 14.1479 /
\plot 10.096 14.1479 10.625 14.1013 /
\plot 10.625 14.1013 11.144 14.0653 /
\plot 11.144 14.0653 11.647 14.0382 /
\plot 11.647 14.0382 12.128 14.0187 /
\plot 12.128 14.0187 12.581 14.0059 /
\plot 12.581 14.0059 13. 14. /
\linethickness=0pt
\putrectangle corners at 3.8 13.8 and 13.6 19.2
\endpicture
\vskip 20pt
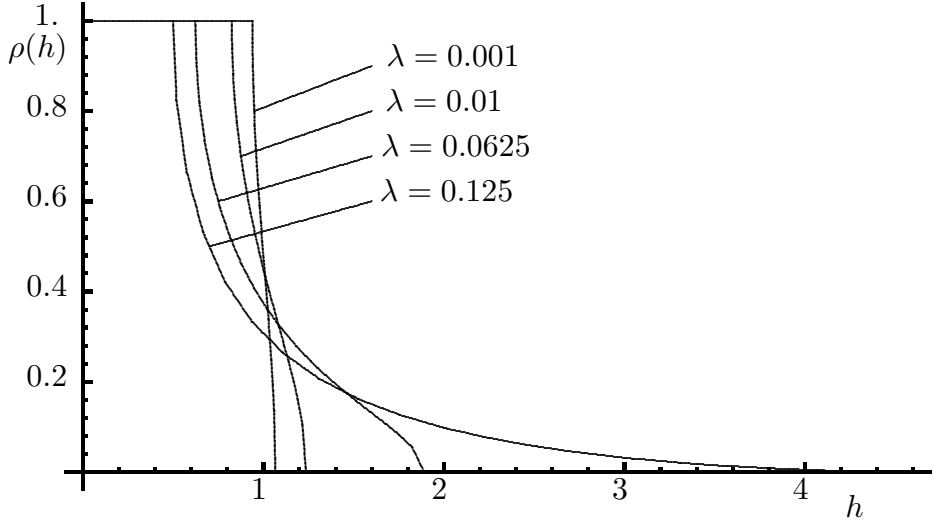
\centerline{{\bf Fig. 4:} Highest weight density $\rho(h)$ for
potential $-\ln(1-M^2)$}
\vskip 20pt
It is easy to check the correctness of this solution by perturbation
theory.  Even better, we can independently calculate the eigenvalue
resolvent $W(P)$ of this model by traditional methods and demonstrate
its exact coincidence with the resolvent obtained from the inversion
formulae
\resolvinvers.

To demonstrate the power of our method we will next consider a case
that is not a traditional one matrix model and thus has not yet been
solved with other methods: Consider $A=\J$ and $B=\J$, i.e. planar graphs
with even coordination numbers for vertices {\it and} faces.
Here eqs. \IzDiFr\ and \chars\ lead to
\eqn\evev{ Z\sim \sum_{\{h^e,h^o\}}
\prod_i(h^e_i-1)!!h^o_i!!\
{\Delta(h^o)^2\ \Delta(h^e)^2 \over \prod_{ij}(h_i^o - h_j^e)}
\bigl({\lambda\over N}\bigr)^{-{1\over 4}N(N-1)+{1\over
2}\sum_i(h^e_i+h^o_i)}.}
Here even and odd weights no longer factorize. However, it is natural
to assume that they are equally distributed. Thus the crossproduct
should precisely cancel one power of a Vandermonde, leading to
the saddlepoint equation
\eqn\msadevev{\barint_b^a\ dh'\ {\rho(h') \over h-h'} =
- \ln\ ( h\ \lambda) - \ln ({h\over h-b})}
This equation is solved exactly as the previous case and one finds
the weight resolvent
\eqn\Hevev{H(h)=
\ln\left[{(a-b)\over h\lambda(\sqrt{a}+\sqrt{b})^2}
{2h^2-(\sqrt{a}-\sqrt{b})^2+2ab+2(h+\sqrt{ab})\sqrt{(h-a)(h-b)}\over
(a+b)h-2ab+2\sqrt{ab}\sqrt{(h-a)(h-b)}}\right],}
with the interval boundaries being determined through the quantities
$\xi = ({\sqrt{a} + \sqrt{b}\over 4})^2$ and
$\eta = ({\sqrt{a} - \sqrt{b}\over 4})^2$ by
\eqn\abevev{3\lambda^2\xi^3-\xi+1 =0
{\rm \hskip 20pt and \hskip 20pt}
\eta = {1 \over 3}(\xi - 1)}
One easily finds the critical coupling to be $\lambda_c = {2 \over
9}$.  It is satisfying to observe that this is very slightly less than
two times the value of the previous case; this is as expected since
the asymptotic growth of the number of graphs with $n$ edges is $\sim
\lambda_c^{-n}$. It is straightforward to verify that this
solution indeed correctly counts the graphs under consideration; e.g.
from equation \Hevev\ with the help of \resolvents :
$$
\hskip -120pt {1 \over N}\ \Tr M^2 = \lambda + \lambda^3 + 6 \lambda^5 +
54 \lambda^7 + \ldots$$
\eqn\pert{=
\beginpicture
\setcoordinatesystem units < 1.000cm, 1.000cm>
\linethickness=1pt
\plot  2.826 16.605 3.810 16.605 /
\put {\sixrm 1} [lB] at  3.937 16.605
\linethickness=0pt
\putrectangle corners at  2.826 16.732 and  4.000 16.605
\endpicture
\quad\lambda\,\, +\,\,
\beginpicture
\setcoordinatesystem units < 1.000cm, 1.000cm>
\linethickness=1pt
\put {$\bullet$} [1B] at  2.032 16.362
\put {$\bullet$} [1B] at  2.350 16.362
\plot  1.715 16.447  2.667 16.447 /
\put {\sixrm 1} [lB] at  2.794 16.447
\linethickness=0pt
\putrectangle corners at  1.715 16.574 and  2.857 16.415
\endpicture
\quad\lambda^3\quad+\quad
\beginpicture
\setcoordinatesystem units < 1.000cm, 1.000cm>
\linethickness=1pt
\put {$\bullet$} [1B] at  2.381 16.393
\put {$\bullet$} [1B] at  2.667 16.393
\ellipticalarc axes ratio  0.159:0.159  360 degrees
	from  2.540 16.669 center at  2.381 16.669
\put {$\bullet$} [1B] at  2.381 16.711
\put {$\bullet$} [1B] at  2.254 15.854
\put {$\bullet$} [1B] at  2.762 15.854
\ellipticalarc axes ratio  0.254:0.159  360 degrees
	from  2.762 15.939 center at  2.508 15.939
\put {$\bullet$} [1B] at  2.667 16.965
\put {$\bullet$} [1B] at  2.826 16.965
\put {$\bullet$} [1B] at  2.191 16.965
\put {$\bullet$} [1B] at  2.413 16.965
\plot  2.032 16.478  3.016 16.478 /
\plot  2.032 15.939  3.016 15.939 /
\plot  2.032 17.050  3.016 17.050 /
\put {\sixrm 1} [lB] at  3.112 16.986
\put {\sixrm 4} [lB] at  3.112 16.510
\put {\sixrm 1} [lB] at  3.112 15.939
\linethickness=0pt
\putrectangle corners at  2.032 17.113 and  3.175 15.780
\endpicture
\quad\lambda^5\quad+\quad
\beginpicture
\setcoordinatesystem units < 1.000cm, 1.000cm>
\linethickness=1pt
\circulararc 87.206 degrees from  3.429 17.018 center at  3.588 17.185
\circulararc 180.000 degrees from  3.429 17.018 center at  3.588 17.018
\circulararc 135.675 degrees from  3.429 16.192 center at  3.487 16.050
\circulararc 96.733 degrees from  3.429 15.907 center at  3.302 16.050
\circulararc 180.000 degrees from  3.873 15.399 center at  3.588 15.399
\circulararc 95.850 degrees from  3.302 15.399 center at  3.588 15.657
\circulararc 113.767 degrees from  3.524 15.589 center at  3.475 15.422
\circulararc 111.233 degrees from  3.302 15.399 center at  3.375 15.565
\put {$\bullet$} [1B] at  1.524 17.631
\put {$\bullet$} [1B] at  2.254 17.631
\put {$\bullet$} [1B] at  1.683 17.631
\put {$\bullet$} [1B] at  2.127 17.631
\put {$\bullet$} [1B] at  1.841 17.631
\put {$\bullet$} [1B] at  1.969 17.631
\put {$\bullet$} [1B] at  1.715 14.711
\put {$\bullet$} [1B] at  2.032 14.711
\put {$\bullet$} [1B] at  1.651 14.584
\put {$\bullet$} [1B] at  2.127 14.584
\ellipticalarc axes ratio  0.254:0.159  360 degrees
	from  2.127 14.669 center at  1.873 14.669
\put {$\bullet$} [1B] at  2.064 14.965
\put {$\bullet$} [1B] at  2.254 14.965
\put {$\bullet$} [1B] at  1.532 14.965
\put {$\bullet$} [1B] at  1.841 14.965
\ellipticalarc axes ratio  0.159:0.159  360 degrees
	from  1.841 15.050 center at  1.683 15.050
\put {$\bullet$} [1B] at  1.715 15.473
\put {$\bullet$} [1B] at  2.064 15.473
\ellipticalarc axes ratio  0.159:0.159  360 degrees
	from  1.873 15.716 center at  1.715 15.716
\ellipticalarc axes ratio  0.095:0.095  360 degrees
	from  1.810 15.780 center at  1.715 15.780
\put {$\bullet$} [1B] at  1.715 15.630
\put {$\bullet$} [1B] at  1.715 15.790
\put {$\bullet$} [1B] at  1.587 16.107
\put {$\bullet$} [1B] at  1.873 16.107
\put {$\bullet$} [1B] at  1.715 16.012
\put {$\bullet$} [1B] at  2.064 16.012
\ellipticalarc axes ratio  0.159:0.159  360 degrees
	from  1.873 16.192 center at  1.715 16.252
\put {$\bullet$} [1B] at  1.715 16.298
\put {$\bullet$} [1B] at  1.683 16.520
\ellipticalarc axes ratio  0.159:0.159  360 degrees
	from  1.841 16.796 center at  1.683 16.796
\put {$\bullet$} [1B] at  1.683 16.838
\put {$\bullet$} [1B] at  2.095 16.520
\ellipticalarc axes ratio  0.159:0.159  360 degrees
	from  2.254 16.796 center at  2.095 16.796
\put {$\bullet$} [1B] at  2.095 16.838
\put {$\bullet$} [1B] at  2.159 17.092
\put {$\bullet$} [1B] at  1.969 17.092
\put {$\bullet$} [1B] at  1.587 17.092
\put {$\bullet$} [1B] at  1.746 17.092
\ellipticalarc axes ratio  0.159:0.159  360 degrees
	from  1.905 17.335 center at  1.746 17.335
\put {$\bullet$} [1B] at  1.746 17.377
\put {$\bullet$} [1B] at  3.365 17.441
\put {$\bullet$} [1B] at  3.747 17.441
\ellipticalarc axes ratio  0.159:0.095  360 degrees
	from  3.524 17.621 center at  3.365 17.621
\ellipticalarc axes ratio  0.159:0.095  360 degrees
	from  3.524 17.462 center at  3.365 17.462
\put {$\bullet$} [1B] at  3.365 17.631
\put {$\bullet$} [1B] at  3.365 17.314
\put {$\bullet$} [1B] at  3.429 16.933
\put {$\bullet$} [1B] at  3.747 16.933
\ellipticalarc axes ratio  0.159:0.159  360 degrees
	from  3.905 17.177 center at  3.747 17.177
\put {$\bullet$} [1B] at  3.747 17.250
\put {$\bullet$} [1B] at  3.429 16.393
\put {$\bullet$} [1B] at  3.747 16.393
\ellipticalarc axes ratio  0.159:0.159  360 degrees
	from  3.588 16.605 center at  3.429 16.605
\put {$\bullet$} [1B] at  3.429 16.647
\ellipticalarc axes ratio  0.095:0.095  360 degrees
	from  3.524 16.542 center at  3.429 16.542
\put {$\bullet$} [1B] at  3.429 16.520
\put {$\bullet$} [1B] at  3.429 15.822
\put {$\bullet$} [1B] at  3.747 15.822
\ellipticalarc axes ratio  0.159:0.159  360 degrees
	from  3.588 16.066 center at  3.429 16.066
\put {$\bullet$} [1B] at  3.429 16.107
\put {$\bullet$} [1B] at  3.302 15.314
\put {$\bullet$} [1B] at  3.873 15.314
\put {$\bullet$} [1B] at  3.556 15.504
\put {$\bullet$} [1B] at  3.302 14.837
\put {$\bullet$} [1B] at  3.873 14.837
\ellipticalarc axes ratio  0.254:0.095  360 degrees
	from  3.842 14.922 center at  3.588 14.922
\ellipticalarc axes ratio  0.254:0.191  360 degrees
	from  3.842 14.922 center at  3.588 14.922
\plot  1.429 17.716  2.350 17.716 /
\plot  1.429 14.669  2.350 14.669 /
\plot  1.429 15.050  2.350 15.050 /
\plot  1.429 15.558  2.350 15.558 /
\plot  1.429 16.097  2.350 16.097 /
\plot  1.429 16.605  2.350 16.605 /
\plot  1.429 17.177  2.350 17.177 /
\plot  1.429 17.177  2.350 17.177 /
\plot  3.080 17.526  4.064 17.526 /
\plot  3.080 17.018  4.064 17.018 /
\plot  3.080 16.478  4.064 16.478 /
\plot  3.080 15.907  4.064 15.907 /
\plot  3.080 15.399  4.064 15.399 /
\plot  3.080 14.922  4.064 14.922 /
\put {\sixrm 1} [lB] at  2.508 17.653
\put {\sixrm 8} [lB] at  2.508 17.177
\put {\sixrm 4} [lB] at  2.508 16.669
\put {\sixrm 4} [lB] at  2.508 16.097
\put {\sixrm 8} [lB] at  2.508 15.526
\put {\sixrm 3} [lB] at  2.508 15.081
\put {\sixrm 3} [lB] at  2.508 14.669
\put {\sixrm 6} [lB] at  4.159 17.526
\put {\sixrm 6} [lB] at  4.159 17.050
\put {\sixrm 4} [lB] at  4.159 16.415
\put {\sixrm 4} [lB] at  4.159 15.907
\put {\sixrm 2} [lB] at  4.159 15.399
\put {\sixrm 1} [lB] at  4.159 14.891
\linethickness=0pt
\putrectangle corners at  1.429 17.780 and  4.223 14.510
\endpicture
\quad\lambda^7\quad+\,\,\dots}
where the dots correspond to insertions of the matrix $\J$.

These examples correspond to the ensembles of planar graphs simplest
in the weight language. It is natural to ask for the description
of the simplest original even model \BIPZ\ of pure gravity, i.e.~the
one matrix model with the action
$-{1\over 2}\Tr M^2+{\lambda \over 4} \Tr M^4$. It is
again straightforward to explicitly calculate the characters here
(see Appendix);
the weights are now grouped into four blocks $h_i^{(\epsilon)}$
where $\epsilon \in \{0,1,2,3\}$ denotes their congruence, modulo four.
This leads to the expansion
\eqn\mfour{\eqalign{Z\sim
\lambda^{-{N^2\over 8}}
&\sum_{\{h^0,h^2\}}
\D(h^{(0)})^2\ \D(h^{(2)})^2\ \prod_{i,j}(h^{(2)}_i-h^{(0)}_j)\,
e^{\Sigma_k^{\epsilon=0,2}\,{h_k^{(\epsilon)}\over 4}
(\ln(\lambda h_k^{(\epsilon)}/N)-1)}\cr
&\sum_{\{h^1,h^3\}}
\D(h^{(1)})^2\ \D(h^{(3)})^2\ \prod_{i,j}(h^{(3)}_i-h^{(1)}_j)\,
e^{\Sigma_k^{\epsilon=1,3}\,{h_k^{(\epsilon)}\over 4}
(\ln(\lambda h_k^{(\epsilon)}/N)-1)},}}
where for convenience we have substituted in Sterling's
formula for the factorials.
One observes that even and odd weights factorize, but not
the congruence classes $(0,2)$ and $(1,3)$.
In fact, each of the non-factorizing sectors has a structure identical
to the case $A=B=1$ (i.e. the one matrix
model with action $-{1\over 2\lambda}\Tr M^2- \ln (1-M)$)
since here the character expansion \IzDiFr\ gives, together with \chars,
\eqn\badlog{Z \sim
\lambda^{-{N^2\over 4}}
\sum_{\{h^e,h^o\}}
\Delta(h^o)^2\ \Delta(h^e)^2\ \prod_{ij}(h_i^o - h_j^e)\
e^{\Sigma_k\,{h_k\over 2}(\ln(\lambda h_k/N)-1)}.}
Here it is the even and odd weights that remain coupled.  We see
directly that the partition function in equation \mfour\ is the square
of the partition function in equation \badlog . We thus rediscover in
the highest weight language the well known connection between these
two models. At the diagrammatic level this can be seen by placing a
vertex of the $M^4$ model at the midpoint of every edge of the $-\ln
(1-M^2)$ diagrams so that the face centres of the $M^4$ model are the
vertices and face centres of the $-\ln (1-M^2)$ model.  It is tempting
to make a saddlepoint ansatz like in \msadevlog, multiplying in this
equation the principal part integral by an extra factor of ${3 \over
2}$ due to the variation of the extra factor $\prod_{ij}(h_i^o -
h_j^e)$.  However, here the solution of this equation does {\it not}
lead to the correct result. We can gain some insight into this failure
by computing, by the usual means, the eigenvalue resolvent $W(P)$ and
then deducing $H(h)$ from \resolvinvers. The result of this
calculation leads to a third order algebraic equation for $e^{H(h)}$:
\eqn\mforesolv{
H(h)=\ln({X(h)\over h})}
with $X(h)$ defined through the solution of
\eqn\cardano{\lambda X^3-\lambda(1+h)X^2+
({8\over 9}-h+\gamma(\lambda))X+h^2=0}
with $\gamma(\lambda)=
{1\over 54}{(1-\sqrt{1-12\lambda})(1-12\lambda)\over 12\lambda}$.
One then finds on
inspecting $H(h)$ that the saddlepoint configuration of weights
is {\it complex}: The rapid sign-changes of the product
$\prod_{ij}(h_i^o - h_j^e)$ destabilize the reality of the saddlepoint.
It is worth pointing out that the saddle point nevertheless
{\it exists}, even though it is much harder to find.
As we have seen in the previous example, the presence of
this factor in the denominator of the expansion is however without
danger. A rough intuitive ``explanation''  is that in the numerator
the product acts to repulse the different distributions,
destabilizing the saddle point, whereas in the
denominator it attracts and stabilizes.
We will see in the next section that the stability of the saddlepoint
can be preserved in the case of greatest physical interest:
the gradual flattening of the random surface.

\newsec{Flattening random lattices}

Before we flatten our surface it is worth understanding how the flat
lattice is represented in the language of highest weights. In this
case ${1\over N}\Tr\ A^q={1\over N}\Tr\ B^q=\delta_{q,4}$ and it is
simple to derive the characters from equation \schurchar\ (see
Appendix), to obtain the partition function
\eqn\flatZ{
Z\sim\lambda^{-{N^2\over 4}}\sum_{\{h^0,h^1,h^2,h^3\}}
{\D(h^{(0)})^2\D(h^{(1)})^2\D(h^{(2)})^2\D(h^{(3)})^2\quad
e^{\Sigma_{k,\epsilon}\,{1\over 2}h^{(\epsilon)}_k\ln(\lambda)}\over
\prod_{i,j}(h^{(1)}_i-h^{(0)}_j)(h^{(3)}_i-h^{(0)}_j)
(h^{(1)}_i-h^{(2)}_j)(h^{(3)}_i-h^{(2)}_j)},}
where again we have substituted in Sterling's formula for the
factorials.  The potential term,
$e^{\Sigma_{k,\epsilon}\,{1\over 2}h^{(\epsilon)}_k\ln(\lambda)}$,
attracts to the origin for $\lambda<1$, and repulses and is unstable
for $\lambda>1$.  The critical point is therefore now $\lambda_c=1$.
The repulsion of the Vandermondes in the numerator is now precisely
balanced by the attractive effect of the products in the denominator.
Indeed, using our rule of thumb that the variation of a product
$\prod_{i,j}(h^{(\epsilon_2)}_i-h^{(\epsilon_1)}_j)$ in the
denominator precisely cancels the variation of one power of a
Vandermonde, we arrive at the trivial saddle point equation
\eqn\fltsdl{-{1\over 2}\ln(\lambda)=0.}
We observe that the saddle point equation is not satisfied anywhere
(except possibly for $\lambda_c$) i.e. none of the weights are excited
and, as in the gaussian case, only the empty Young tableau
contributes:
\eqn\fltH{H(h)=\ln{h\over h-1},}
To order $N^2$, only the original gaussian term can contribute to the
partition function and expectation values of $\Tr M^q$. The potential
term cannot contribute to either, since it is impossible for all loops
to have 4 $A$ matrices running around them. This is simply the
statement of the fact that it is impossible to put a completely flat
lattice on the sphere. Positive curvature defects have to be
introduced to close the surface.

Since the order $N^2$ contribution is trivially zero, it is
interesting to investigate the first non-trivial order: the order
$N^0$ contribution.  We will give some qualitative arguments that the
flat case is described by N fermions with an equidistant spectrum and,
from this starting point, calculate the number of flat graphs in the
first 2 orders in $1/N$.

One may notice that, in the large N limit, all the factorials and
products of differences of various highest weights cancel (assuming
all 4 groups of $h$'s to be distributed in the same way). One can
hypothesize that this will be true even for the next order in $1/N^2$
which describes the graphs with the topology of a torus (we will check
this assumption below). We are then left with the partition function
of fermionic type:
\eqn\Zferm{
Z(\lambda)=\lambda^{N(N-1)/2}\sum_{h_1>h_2...>h_N} \lambda^{\Sigma h}.}
A standard calculation for the free energy $f(\lambda)=1/N^2
\ln Z(\lambda)$
gives:
\eqn\fflat{
f(\lambda) = -\sum_{n=1}^{N} \ln(1-\lambda^n)
= - N^{-2} \sum_{n=1}^{\infty}\ln(1-\lambda^n)+ O(N^{-4})
= N^{-2} \lambda^{-1/24} \eta(\lambda) + O(N^{-4}),}
where $\eta(\lambda)$ is the Dedekind function.  The order $N^2$
contribution, which counts the number of flat graphs with spherical
topology, is zero since, as already stated above, no such graphs
exist, at least not without defects. The next order counts the number
of flat graphs that can be fitted onto a torus. It is easy to verify
this calculation by directly counting the number of possible graphs. A
general graph on the torus consists of $m\times n$ squares glued into
a rectangle ($m$ columns and $n$ rows). Opposite sides are then glued
together: first we glue together the two sides of the length $n$ and
then, with $m$ possible twists, the two sides of length $m$. The
symmetry factor is $1/(mn)$. We obtain:
\eqn\flator{
f(\lambda) = \sum_{n,m=0}^{\infty} {m\over mn}\lambda^{mn} =
-\sum_{n=0}^{\infty} \ln(1-\lambda^n),}
which coincides exactly with our result obtained from the sum over
highest weights treated as free fermions.
Since  completely flat graphs exist only for the torus topology we can
be sure that there exist no higher order contributions.
Let us note that this calculation on the torus is similar to what
we would have for QCD2 on a toroidal target space \GrossTaylor\  .

We conclude that, in the limit of flat graphs, the highest weights
play the role of the energy levels of N free fermions.
This fact may be useful for the understanding of the mechanism of
the flattening phase transition.

We are now in a position to discuss the real problem of interest. We
have observed that, for models \evev\ and \flatZ , where both the
vertex and face coordination numbers are even, the simple saddle point
equation is valid. In fact, all the simple models with even face and
vertex coordination numbers which we have investigated (for brevity we do
not discuss them here) can be solved correctly with the simple saddle
point formulation. We thus hypothesize that if we restrict our attention to the
``even-even'' models (models where the coordination numbers for both
faces and vertices are even), the destabilization of the saddle point
discussed at the end of section 5 will be avoided and we can correctly
interpolate all the way from the $A=B=\J$ case \evev\ to the flat
lattice \flatZ\ using the saddle point approximation. Any results we
obtain can be checked against perturbation theory to verify this
assumption.  For the interpolating model we thus choose
\eqn\ABev{A=B=\left[\matrix{\sqrt{b}&0\cr 0&-\sqrt{b}\cr}\right],}
so that all odd traces equal zero, calculate the character from
\eigchar\   (see Appendix)
\eqn\ABevchar{
\chi_{\{h\}}(A)=\chi_{\{h\}}(B)=
    \chi_{\{{h^e\over 2}\}}(b)\chi_{\{{h^o-1\over 2}\}}(b)
    \,\,\sgn\bigl[\prod_{i,j}(h^e_i-h^o_j)\bigr],}
and arrive at the partition function
\eqn\RsqZ{\eqalign{Z\sim \sum_{\{h^e,h^o\}}
\prod_i(h^e_i-1)!!h^o_i!!\
{\Delta(h^o)^2\ \Delta(h^e)^2 \over \prod_{ij}(h_i^o - h_j^e)}
&\bigl({\lambda\over N}\bigr)^{{1\over 4}N(N+1)+{1\over 2}\sum_i(h^e_i+h^o_i)}
\cr&\times\,\,I({h^e\over 2},b)^2\,I({h^o-1\over 2},b)^2,}}
where
$I({h^e\over 2},b)=\chi_{\{{h^e\over 2}\}}(b)\D(b)/(\D(\beta)\D({h^e\over 2}))$
is an Itzykson Zuber integral, with $b=e^{\beta}$ and
$I({h^o-1\over 2},b)$ is defined similarly.  The partition function is a
generalisation of \evev\ the difference being the two Itzykson Zuber
integrals.  Indeed, setting $b=1$ we recover \evev .  For convenience
we introduce the notation $\tilde t_q={2\over qN}\Tr\ b^q$ so that
$\tilde t_q=t_{2q}=t_{2q}^*$.  In this notation, flattening of the
lattice corresponds to setting $\tilde t_2\neq0$ and $\tilde t_q=0$
for $q\neq2$. In fact, in complete analogy with the ``rainbow''
$\longrightarrow$ ``cigar-like'' transition in the gaussian model, it
is only necessary to set $\tilde t_q=0$ for $q>2$ to flatten the
lattice. For convenience, we define $F(h^e_k)=2{\partial\over\partial
h^e_k}I({h^e\over 2},b)$. It is then easy to derive the following pair
of equations which in principle completely describe the model:
\eqn\Trb{q\tilde t_q={2\over N}\Tr\ b^q={1\over q}\oint\,{dh\over 2\pi i}
e^{q(H(h)+F(h))},}
\eqn\Rsqsdpt{\barint_b^a\ dh'\ {\rho(h') \over h-h'} =
-\ln(\lambda h)-\ln{h\over h-b}-2F(h).}
The first comes from a generalisation of \trchar\  and the second is
the saddle point equation. The whole complexity of the random to flat
transition is succinctly encapsulated in these two equations.
As in the toy, gaussian model, we can capture the transition by
turning on the first three coupling constants $\tilde t_q$. However, this
system of equations, though simpler than the original statistical
mechanics problem \DWG , is still highly non-trivial. As it
is a kind of Riemann-Cauchy problem, the difficulty
is as always to guess the analytical structure of the solution.

\newsec{Conclusions and discussion}

In this paper, we developed a character expansion technique for a new
kind of matrix model describing dually weighted planar graphs. We first
reduced the $N^2$ degrees of freedom of the original matrix to the $N$
highest weights specifying the irreducible representations of $U(N)$.
This allowed us, with some precautions, to apply the saddle point
method in the large N limit, reducing the problem to a set of integral
equations.

We showed how to solve these integral equations in a number of known
1-matrix models and also found a new result: we are able to calculate
the number of graphs having only even numbers of neighbours for both
original and dual vertices.

We also understood the limiting case of the completely flat lattice.
It is described by a system of free fermions (the highest weights) with
an equidistant spectrum. This allowed us to reproduce correctly the
partition function of regular graphs with toric topology (the only
genus that can be realized from a completely flat graph).

The most important physical question still to be addressed is the description
of the transition from completely random planar graphs
(describing the 2D gravity) to the regular (flat, in our terminology)
lattice.

We have not yet solved the corresponding integral equations. It is not
an easy task as they are equivalent to a complicated Cauchy-Riemann
problem, and the solution involves some non trivial guesses
about the analytical structure of the underlying functions.

Let us speculate possible physical pictures for the flattening phase
transition. There are three different scenarios to consider:

1. The flattening could take place for a finite effective coupling
constant in front of the $R^2$ (curvature squared) term in the 2D
gravity action.  This means that the characteristic flat size of an
almost flat piece of a graph diverges at some finite critical
coupling. This would be the most interesting scenario as it would mean
the discovery of a completely new universal critical phenomenon.

An argument against this picture is the fact that the $R^2$
coupling is dimensionfull, thus containing inverse powers of the
cut-off. This is however a completely perturbative argument since the
same reasoning could be applied to the 3D Ising model described as
$\phi^4$ scalar field theory. Here, the interaction term is also
dimensionfull, but it nevertheless exhibits a non-perturbative phase
transition.

A more serious objection to a transition at finite effective coupling
is the presence of macroscopic excitations on the background of
a regular lattice generated by only a very small number of
curvature defects. For example, on the square lattice, the
introduction of four vertices of coordination number 6 and four of
coordination number 2 allows a baby universe of arbitrary size to grow
out from the flat lattice. This is in complete analogy with the tree-like
structures studied in the toy gaussian model of section 4, where the
introduction of just one $t^*_3$ defect and one $t^*_1$ defect is
enough to create a whole new branch.

2. The flattening could appear only in the limit of the infinite
$R^2$ coupling. In the light of the previous arguments this
is the most likely scenario. Another argument was given in
\Kawai\  using the methods of Liouville theory, though again the
argument was completely perturbative and thus unsatisfactory.
This hypothesis also seems to be in agreement with numerical
simulations \BK\  .
Even if this scenario turns out to be correct, the asymptotic approach
to the flat lattice could contain some interesting scaling behaviour
and is worthy of study.

3. The third scenario is an extended version of the second one: we
could have a flattening phenomenon for the $R^2$ coupling of the order
of $\Lambda_{cutoff}^2$. It will of course depend upon the
type of regular lattice, and the critical exponents, if any, will be
dependent on the symmetry of the lattice; triangular or square. In
this case the phase transition would be better identified as a
crystalization transition. The universal critical properties would then
depend on the symmetry of the "crystal".

One could think of an analogy with the Berezinski-Kosterlitz-Thouless
phase transition: the discrete curvatures on the lattice could be
identified with quantized coulomb charges in two dimensions. This is
indeed the picture in the conformal gauge of the continuous 2D
gravity action.

Let us also make some technical comments.
The character expansion method can be successfully applied to a more
general type of matrix model, having the following action:
\eqn\genact{
S(M)= tr( W(M) + V(AM) )}
instead of just $W(M)=M^2$ as in the present paper. The character
expansion coefficients are more tricky, as in \IDiF\ ,  but
there are no real obstacles: we still deal with only N
highest weights as the principal degrees of freedom.

One can imagine the solution of a generalized 2-matrix model,
such as
\eqn\dream{
Z=\int d^{N^2}L\,\, d^{N^2}M\,\, \exp\left(\Tr( V(L)+V(M)+W(LM) )\right),}
with an arbitrary function $W(x)$, not just a merely linear one, or even a
similar multi-matrix chain.

The most ambitious step would be to put "matter" on a random, but
gradually flattening, lattice. For example, one can consider the
matrix action:
\eqn\flatmat{
S(M)= tr( W(AL) + V(AM) + L^2+ M^2 + c LM ).}
It describes the Ising model on DWG. This model provides the
interpolation between two solvable cases: the Onsager solution for a
regular lattice and the Ising model on dynamical random graphs
\Kaz\ . If $W \neq V$ it includes the non-zero magnetic field,
which is still an unsolved case for the regular lattice.
Unfortunately, the character expansion would be much more complicated
object in this case as it would contain some non-trivial
Klebsch-Gordan coefficients.

In any case, we believe that the proposed approach could be fruitful
for attacking many new combinatorial problems in 2D statistical
mechanics and field theory.

\newsec{Appendix}
Below we present explicit formulea for some characters derived
from the definitions in section 2.
\subsec{\bf $A = 1${\rm, the unit matrix}}
In this case the character is just the dimension of the representation.
The easiest way to derive this is to take the limit as
$\epsilon\rightarrow 0$ of the character formula,
equation \eigchar\  with $a_k=e^{k\epsilon}$. In this case
\eqn\charone{
\chi_{\{h\}}(1)=\lim_{\epsilon\to 0}
 {\Delta((e^{\epsilon h_l})^k)\over \Delta(e^{\epsilon k})}
               =c\,\,\Delta(h_i),}
where $c$ is the numerical constant $c=\prod_{i=1}^{N-1}i!$ .
\subsec{\bf $A_m${\rm, defined by }${1\over N}\Tr (A_m^k)=\delta_{k,m}$ }
The traces of all positive powers of $A_m$ are zero except $(A_m)^m$.
We will sketch the derivation for the case $m=2$. It is easy to
generalize the derivation for arbitrary $m$.  Using the second
definition for the character, equations \schurpolyn\  \schurchar , we
have
\eqn\PAtwo{
P_k=\cases{
   {1\over (k/2)!}\bigl({N\over 2}\bigr)^{k/2}\quad &
                               $k$ even and non-negative;\cr
   0 & otherwise}.}
If we substitute this into the determinant, we obtain a matrix
structure in which every other entry in a row is zero. By
interchanging rows and columns, the determinant can be put into block
diagonal form, with one block for the $h^e$'s and the other for the
$h^o$'s. The powers of $N/2$ factor out and if we
then factor out the product $\bigl(\prod_i({h^e_i\over
2})!({h^o_i-1\over 2})!\bigr)^{-1}$, the entries in the diagonal blocks
become polynomials of ascending order in $h^e$ or $h^o$, and the
block determinants reduce to Vandermonde determinants. Taking into
account all of the sign changes from reordering the rows and columns we
obtain:
\eqn\chitwo{
\chi_{\{h\}}(A_2)=
c\,\,\left({N\over 2}\right)^{{1\over 2}\Sigma_ih_i}
{\Delta(h^e)\Delta(h^o)\over
\prod_i\bigl({h^e_i\over 2}\bigr)!\bigl({h^o_i-1\over 2}\bigr)!}
\,\,\sgn\bigl[\prod_{i,j}(h^e_i-h^o_j)\bigr],}
where $c$ is a numerical constant. For general $m$ we have:
\eqn\chim{
\chi_{\{h\}}(A_m)=
c\,\,\left({N\over m}\right)^{{1\over m}\Sigma_ih_i}\quad
\prod_{\epsilon=0}^{m-1}{\Delta(h^{(\epsilon)})\over
\prod_i\bigl({h^{(\epsilon)}_i-\epsilon\over m}\bigr)!}
\,\,\sgn\bigl[\prod_{0\leq\epsilon_1<\epsilon_2\leq(m-1)}
\prod_{i,j}(h^{(\epsilon_2)}_i-h^{(\epsilon_1)}_j)\bigr].}
In this case, the integers $h$ factor into $m$ groups of $N\over m$
integers $h^{(\epsilon)}$ with $\epsilon=0,1,\dots,(m-1)$ denoting their
congruence modulo $m$.
\subsec{$A_{bb}$ defined by $\Tr\ (A_{bb})^q=0$ for odd $q$ }
Only the even powers of $A_{bb}$ are non-zero. The matrix $A_{bb}$ can
be defined by an ${N\over 2}$ by ${N\over 2}$ matrix, $b$,
(eigenvalues $b_k$) as follows:
\eqn\evA{
A_{bb}=\left[\matrix{b&0\cr 0&-b\cr}\right].}
Again we just sketch the derivation. Calculating the determinant in
equation \eigchar\  and rearranging the columns we notice that we can
write it (up to some sign factors from the interchanges) as
\eqn\bbdet{
det_{_{\hskip -2pt (k,l)}}(a_k^{h_l})=
           \left|\matrix{b^{h^e}&b^{h^o}\cr
                         b^{h^e}&-b^{h^o}\cr}\right|
                                     =
           (-2)^{N\over 2}\left|\matrix{b^{h^e}&0\cr
                         0&b^{h^o}\cr}\right|,}
where, for notational convenience, we denote by $b^{h^e}$ the $N\over 2$
by $N\over 2$ matrix whose elements are $b_i^{h^e_j}$.
The Vandermonde in equation \eigchar\  is just the special case of the
above result, \bbdet, with $h^e_j=2j-2$ and $h^o_l=2l-1$. Including
the sign factors neglected earlier we arrive at the formula
\eqn\bbchi{
\chi_{\{h\}}(A_{bb})=
    \chi_{\{{h^e\over 2}\}}(b^2)\chi_{\{{h^o-1\over 2}\}}(b^2)
    \,\,\sgn\bigl[\prod_{i,j}(h^e_i-h^o_j)\bigr].}
To obtain the character for the $\J$ matrix introduced in section 5, we
set $b=1$ and obtain the expression given in \chars .

It is easy to generalize to higher order cases. For example, to
study the case where only every third power of the matrix $A_{bbb}$ has a
non-zero trace, we start with an ${N\over 3}$ by ${N\over 3}$ matrix,
$b$, and define
\eqn\cubeA{
A_{bbb}=\left[\matrix{b&0&0\cr 0&\omega b&0\cr 0&0&\omega^2 b\cr}\right],}
where $\omega$ is the third root of unity. This time the character
factors into three characters, one for each of the congruence classes,
modulo three, of $h$.
\vskip 30pt
\hskip -20pt {\bf Acknowledgements}

We would like to thank E. Brezin, J-M. Daul, M. Douglas, A.  Matytsin,
A.  Migdal, I. Kostov, D. Kutasov, and A. Zamolodchikov for many
useful discussions. We would also especially like to thank Ph.
DiFrancesco and C. Itzykson for their enthusiastic interest during the
early stages of this work.

\listrefs
\end